\documentclass[11pt,a4paper]{article}
\pdfoutput=1 
\usepackage{jcappub}
\usepackage{graphicx}
\usepackage{dcolumn}
\usepackage{amsfonts}
\usepackage{amssymb}
\usepackage{bm}
\usepackage{url}
\usepackage{caption}
\usepackage{subcaption}
\usepackage[sort&compress]{natbib}
\usepackage{txfonts}
\usepackage{color}
\usepackage{hyperref}

\begin{document}
\newcommand{\changeR}[1]{\textcolor{red}{#1}}

\newcommand{\TBB}{{{T_{\rm BB}}}}
\newcommand{\TCMB}{{{T_{\rm CMB}}}}
\newcommand{\Te}{{{T_{\rm e}}}}
\newcommand{\Teq}{{{T^{\rm eq}_{\rm e}}}}
\newcommand{\Ti}{{{T_{\rm i}}}}
\newcommand{\nB}{{{n_{\rm B}}}}
\newcommand{\nHe}{{{n_{\rm ^4He}}}}
\newcommand{\nHet}{{{n_{\rm ^3He}}}}
\newcommand{\nHt}{{{n_{\rm { }^3H}}}}
\newcommand{\nHtw}{{{n_{\rm { }^2H}}}}
\newcommand{\nBes}{{{n_{\rm { }^7Be}}}}
\newcommand{\nLis}{{{n_{\rm { }^7Li}}}}
\newcommand{\nLisi}{{{n_{\rm { }^6Li}}}}
\newcommand{\nS}{{{n_{\rm s}}}}
\newcommand{\Teff}{{{T_{\rm eff}}}}

\newcommand{\id}{{{\rm d}}}
\newcommand{\aR}{{{a_{\rm R}}}}
\newcommand{\bR}{{{b_{\rm R}}}}
\newcommand{\neb}{{{n_{\rm eb}}}}
\newcommand{\neql}{{{n_{\rm eq}}}}
\newcommand{\kB}{{{k_{\rm B}}}}
\newcommand{\EB}{{{E_{\rm B}}}}
\newcommand{\zmin}{{{z_{\rm min}}}}
\newcommand{\zmax}{{{z_{\rm max}}}}
\newcommand{\YBEC}{{{Y_{\rm BEC}}}}
\newcommand{\YSZ}{{{Y_{\rm SZ}}}}
\newcommand{\rhob}{{{\rho_{\rm b}}}}
\newcommand{\Ne}{{{n_{\rm e}}}}
\newcommand{\sigT}{{{\sigma_{\rm T}}}}
\newcommand{\me}{{{m_{\rm e}}}}
\newcommand{\nBB}{{{n_{\rm BB}}}}

\newcommand{\hscmb}{{{{\hat{s}_{\rm CMB}}}}}
\newcommand{\scmb}{{{{s_{\rm CMB}}}}}
\newcommand{\wia}{{{{w_{i}^{\alpha}}}}}
\newcommand{\pa}{{{{p^{\alpha}}}}}
\newcommand{\KC}{{{{K_{\rm C}}}}}
\newcommand{\KdC}{{{{K_{\rm dC}}}}}
\newcommand{\Kbr}{{{{K_{\rm br}}}}}
\newcommand{\zdC}{{{{z_{\rm dC}}}}}
\newcommand{\zbr}{{{{z_{\rm br}}}}}
\newcommand{\aC}{{{{a_{\rm C}}}}}
\newcommand{\adC}{{{{a_{\rm dC}}}}}
\newcommand{\abr}{{{{a_{\rm br}}}}}
\newcommand{\gdC}{{{{g_{\rm dC}}}}}
\newcommand{\gbr}{{{{g_{\rm br}}}}}
\newcommand{\gff}{{{{g_{\rm ff}}}}}
\newcommand{\xe}{{{{x_{\rm e}}}}}
\newcommand{\alphafs}{{{{\alpha_{\rm fs}}}}}
\newcommand{\YHe}{{{{Y_{\rm He}}}}}
\newcommand{\pmin}{{{{p_{\rm min}}}}}
\newcommand{\pmax}{{{{p_{\rm max}}}}}
\newcommand{\SE}{{{\dot{{\mathcal{E}}}}}}
\newcommand{\SQ}{{{{{\mathcal{E}}}}}}
\newcommand{\SN}{{\dot{\mathcal{N}}}}
\newcommand{\Sn}{{{\mathcal{N}}}}
\newcommand{\muc}{{{{\mu_{\rm c}}}}}
\newcommand{\xc}{{{{x_{\rm c}}}}}
\newcommand{\xH}{{{{x_{\rm H}}}}}
\newcommand{\mT}{{{{\mathcal{T}}}}}
\newcommand{\Ob}{{{{\Omega_{\rm b}}}}}
\newcommand{\Or}{{{{\Omega_{\rm r}}}}}
\newcommand{\Odm}{{{{\Omega_{\rm dm}}}}}
\newcommand{\mdm}{{{{m_{\rm WIMP}}}}}
\newcommand{\Acmb}{{{{A_{\rm CMB}}}}}
\newcommand{\Ayco}{{{{A_{\rm y/CO}}}}}
\newcommand{\Ad}{{{{A_{\rm dust}}}}}
\newcommand{\Td}{{{{T_{\rm dust}}}}}
\newcommand{\betad}{{{{\beta_{\rm dust}}}}}
\newcommand{\fyco}{{{{f_{\nu}^{\rm y/CO}}}}}
\newcommand{\nudo}{{{{\nu_{0}^{\rm dust}}}}}
\newcommand{\Nside}{{{{N_{\rm side}}}}}

\title{Data driven foreground clustering approach to component separation in
  multifrequency CMB experiments: A new Planck CMB map}

\author[a]{Rishi Khatri}

\affiliation[a]{Department of Theoretical Physics, Tata Institute of 
Fundamental Research,  Homi Bhabha Road, Mumbai 400005, India}
\date{\today}
\emailAdd{khatri@theory.tifr.res.in}
\abstract
{We present a new approach to component separation in multifrequency CMB
  experiments by formulating the problem as that of partitioning the sky
  into pixel clusters such that within each pixel cluster the
  foregrounds have similar spectrum, using only the information available in
  the data. Only spectral information is used for partitioning, allowing
  spatially far away pixels  to belong to the same cluster if their
  foreground properties are close. We then apply a modified internal
  linear combination method to each pixel cluster. Since the foregrounds have
  similar spectrum within each cluster, the number of components required
  to describe the foregrounds is smaller compared to all data taken
  together and simple pixel based ILC algorithm works extremely well. We
  test our algorithm in the full focal plane simulations provided by the
  Planck collaboration. We apply our algorithm to the Planck full mission
  data and compare our CMB maps with the CMB maps released by the Planck
  collaboration. We show that our CMB maps are clean and unbiased on a
  larger fraction of the sky, especially at
  the low Galactic latitudes, compared to publicly available maps released
  by the Planck collaboration.
This is important for constraining beyond the
  simplest $\Lambda$CDM
  cosmological models and study of anomalies. Our  cleaned CMB maps are
  made publicly available for use by the cosmology community.
}

\maketitle
\flushbottom
\section{Introduction}
The Cosmic Microwave Background (CMB) experiments observe the sky in broad frequency bands. The
data thus obtained contains not only the CMB but emission from gas and dust
in our own Galaxy as well as the integrated emission from all the
galaxies since the formation of the first stars. There is a broad window where the
CMB dominates over all other astrophysical emissions in a large part of the sky and the CMB
experiments from the ground as well as space have taken advantage of this. With
the Planck experiment \cite{planck} and the ground based experiments such as
South Pole Telescope (SPT) \cite{spt}, Atacama Cosmology Telescope (ACT)
\cite{act} and BICEP2 \cite{bicep2}, the experimental sensitivity has reached a point where we are
limited not by the detector noise but by the residual foreground contamination in the
data. Multifrequency experiments such as Wilkinson Microwave Anisotropy
Probe (WMAP) \cite{wmap} and Planck allow us to use the fact that the CMB and
the foregrounds have different frequency spectrum to separate the CMB from the
other Galactic and extragalactic components present in the data. However,
in any experiment, we have a limited number of channels available while the
foreground properties, the amplitude as well as the shape of the spectrum
of different physical components, such as thermal and spinning dust emission, line
emission from CO and other molecules and atoms, synchrotron emission etc.,
vary over the sky from pixel to pixel.

If we have a physical parametric model for all the cosmological components
and the
foregrounds, and a sufficient number of frequency channels, we can fit 
for the parameters of the model in each pixel. This is the approach
followed in the Commander \cite{eriksen2006} and Linearized Iterative
Least-squares (LIL) \cite{lil} algorithms. 
However, the measured intensity in every frequency band in 
any given pixel is a superposition of many different sources along the
line of sight. For example, thermal dust emission from many molecular clouds
along the line of sight, which may have different physical properties such
as temperature, and different angular sizes on the sky,  may contribute to the intensity observed in a single pixel.
This makes the description of the foreground emissions by simple
parametric models, at the required accuracy, difficult. These difficulties have been the motivation
towards the development of the so called non-parametric or \emph{blind} component separation
methods which need only the spectrum of the cosmological signal of interest
to be known. These methods, in particular,  do not need any knowledge about the foregrounds except that
their spectrum is different from the signal of interest and some additional
statistical requirements about the independence of components or that the
angular variations of the foregrounds are  uncorrelated with the
signal of interest. These methods such as Spectral Estimation Via
Expectation Maximization (SEVEM) \cite{sevem}, Spectral Matching Independent Component
Analysis (SMICA) \cite{smica2003,smica}, Fast Independent Component
Analysis (FASTICA) \cite{fastica2002}, Needlet Internal Linear Combination
(NILC) \cite{nilc}, scale discretized ILC (SILC) \cite{rpl2016}  or iterative ILC approach \cite{say2017} compute the desired 
cosmological signal (e.g. CMB) map as  a linear combination of the
available different frequency channel sky maps. Different choices of how
we combine the available frequency channel maps and what quantity we
optimize to get the weights at different frequencies gives us different
algorithms. These blind algorithms however work best when the 
 foregrounds can be described by the superposition of a small number of
spectral shapes, smaller than the number of frequency channels
available. As we noticed earlier, the foreground properties, specifically
the spectral shape of the foregrounds, varies over the sky. This means that
we cannot apply a blind algorithm to the full sky but must divide the data
into clusters such that within  each cluster the data can be described by a
superposition of a small number of foreground components. Different
criteria for clustering of the data will lead to different solutions for
the cosmological signal. The above mentioned algorithms differ also in how
they cluster or partition the data, with SEVEM clustering the data in pixel
space into a small number of regions based on the \emph{amplitude} of the
foregrounds, SMICA clusters the data in harmonic space with the weights
 a function of the multipole $\ell$ and NILC clusters the data using
 spherical needlets achieving localization or clustering in broad
multipole bands as well as in real space. In all of these cases, the clustering or partitioning of
the data is done without using the information about the foreground
spectrum  available in the data but is instead motivated more by heuristic
arguments and prior assumptions about the foregrounds. Also, all current algorithms use a
single partitioning of data with some smoothing prescription across the
partitions. It is not clear a priori why a particular partitioning scheme
should be chosen over another and whether there exists  a single optimal
partitioning of the data given our limited knowledge of foregrounds and
limited number of frequency channels available.

The main new feature of our approach that distinguishes it from the
existing algorithms
is how we cluster the data. We
will work in pixel space and use Internal Linear Combination (ILC)
\cite{te1996,tegmark1998,wmapilc, Eriksen2004} as our
component separation method within each cluster. However the basic ideas
about the partitioning of data can be applied in any other basis, 
including the spherical harmonic basis, and any other basic component separation
algorithm.
The following two main ideas underlie our new approach to component
separation:
\begin{enumerate}
\item There is no single optimal partitioning of the data, given our
  limited knowledge of the foregrounds. We must
  therefore explore all possible partitionings probabilistically, subject to the constraint in
  the second point. We will see that this approach automatically blurs the
  boundaries between the partition  and thus  does not require any extra smoothing across
  the partition boundaries. By allowing the partitions or clustering of the
  data to vary we essentially want to take into account the
  uncertainties in our knowledge of the foregrounds.
\item The data should be clustered so that the data within each cluster has
  similar foreground properties, in particular the spectrum, since we will
  be using the spectral information to distinguish the signal of interest from
  other components.
\end{enumerate}

We note that such a problem on \emph{clustering} of data is well suited for
machine learning and in that context  it also goes by the name of
\emph{unsupervised learning}. We will however not go by the machine
learning route but follow a very simple prescription for the clustering of
data with the spectral properties of the foregrounds quantified by  a
single parameter i.e. we will cluster data along a single dimension. We will see that
our simple approach already works very well for Planck data and motivates more
sophisticated machine learning based clustering in more than one dimensions
as well as Bayesian
extensions \cite{vw2016} which we leave for future work. In particular, the
small number of high resolution frequency channels available in Planck do
not allow a more sophisticated clustering approach than a single parameter
one we describe below. A single parameter is however not sufficient to
quantify the differences in shapes of multicomponent foregrounds. Future
experiments such as LiteBIRD \cite{litebird}, Primordial Inflation Explorer
(PIXIE) \cite{pixie}, Cosmic Origins Explorer (CORE) \cite{core} and Probe
of Inflation and Cosmic Origins (PICO) \cite{pico}  would have more than twice
the number of channels available in Planck and would allow implementation
of more accurate
multi-parameter clustering as well as use of machine learning to partition
the data.

In this paper we will be only interested in the CMB to demonstrate our
algorithm but it can  equally well be used for any other component for
which the spectrum is known, such as the Sunyaev-Zeldovich (SZ) effect
\cite{zs1969}. We will explore the SZ effect and other applications in separate
publications. We make the full sky cleaned CMB maps constructed from the
Planck public release 2 (PR2) data release \cite{planckpr2} by our algorithm publicly available.

\section{Data driven foreground clustering (FC)}
We want to construct a measure which can tell us how close or far apart 
two pixels are in terms of their foreground properties other than the amplitude. We will be using
Planck data from 70 GHz to 545 GHz, where the main foreground for the CMB
is dust emission. We construct a very simple measure by subtracting the 100
GHz map from the 545 GHz and 353 GHz maps and then take the ratio of the
resulting maps to get the measure $m(p)$ at each pixel $p$,
\begin{align}
m(p)= \frac{T_{\rm 545 GHz}(p)-T_{\rm 100 GHz}(p)}{T_{\rm 353 GHz}(p)-T_{\rm 100 GHz}(p)},
\end{align}
where $T_{\nu}$ is the map at frequency $\nu$ in units of $K_{\rm CMB}$. The choice of measure is
dictated by the fact that we need two foreground dominated channels from
which we want to subtract out the CMB, estimated from a channel which is
not only relatively cleaner but also does not have too much noise. We note that this measure will
capture how fast the total foreground emission is increasing or decreasing
with frequency and therefore would capture information about the dust
emission or strong synchrotron sources. However, such a measure would miss
components like CO emission which are present in some channels but not
others. Using this measure also limits us to use the channels
  which do not have significant synchrotron emission since it is not
  captured by our measure. In particular we will not be using 30 GHz and 44
GHz channels for this reason. In principle it is possible to construct a
second measure which would capture the synchrotron emission, it will
however complicate the clustering. In this early stage of exploring a new
approach to component separation, we want to keep things as simple as
possible and leave further sophistication to future work. Due to these
reasons we do not expect the clustering to be optimal. By optimal
clustering we mean the clustering where foregrounds in each cluster can be
described by a number of components smaller than the number of frequency
channels available. We will see that such a simple measure works reasonably well for
Planck in particular because  CO emission is not a big contaminant for the
CMB. Our results will motivate exploring  more sophisticated clustering for
Planck as well as future experiments with
higher sensitivity and more frequency channels taking into account the line
emissions in each pixel in addition to the continuum emission. 

\begin{figure}
\resizebox{\hsize}{!}{\includegraphics{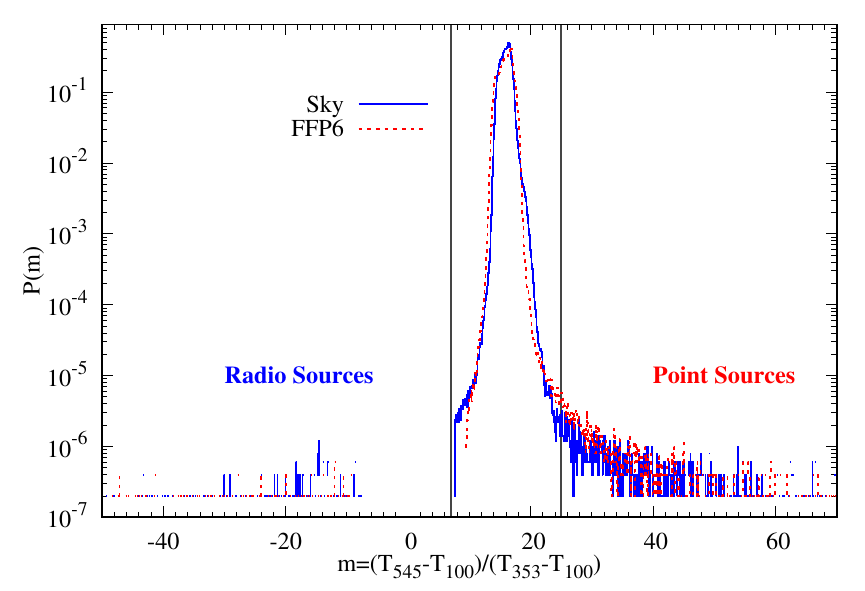}}
\caption{\label{Fig:logmetric}PDF of the measure $m$ for the FFP6
  simulations and Planck PR2 sky. The two tails of the PDF correspond to the
  point sources in radio and far infrared. The two vertical lines show the
  cutoffs we put to remove strong point sources which are put in a separate
  \emph{bad cluster}, cluster number 0 below.}
\end{figure}

To suppress fluctuations in the measure due to the noise, especially from
the 100 GHz channel, and make sure that
we are dominated by signal in every pixel,
 we first smooth all maps to $15'$ before calculating the
measure. In particular we want to suppress noise in the difference maps in
the low foreground pixels at high
latitudes. We show the probability density function (PDF) of the measure, $P(m)$, at  $15'$ resolution, for the Planck FFP6 (Full
Focal Plane)  simulations \cite{ffp6} as well as the PR2 data in Fig. \ref{Fig:logmetric}.
A majority of the pixels are at $7\lesssim m \lesssim 25$ but there is a large tail
at $m>25$ and a smaller tail at $m<0$. The $m>25$ tail is coming from the
infrared point sources while the negative $m$ tail is coming from the low
frequency radio sources for which 353 GHz temperature is smaller than the
100 GHz temperature. The measure $m$  therefore naturally separates out the strongest
point sources and we put all pixels with $m<7$ or $m>25$ into one bad
cluster. We then extend this \emph{bad region} or mask by $5'$. These pixels should be masked in any cosmological analysis. There
are two choices we must now make: How many clusters we should partition the
data into and where should we place the partition boundaries.

\subsection{Choosing the number of pixel clusters (partitions) }
We want to partition the rest of the pixels between our cutoffs, $7\le m
\le 25$, into smaller clusters and apply  the ILC algorithm to each
cluster. The number of clusters should not be too large, since then the ILC
bias in each cluster could be large when we have only a small amount of data
in each cluster. The number of clusters  should not be too small since then the foregrounds
in each cluster would be too complicated. We choose to have 12 clusters + 1
bad cluster of point sources giving a total of 13 clusters of pixels. Our
algorithm is not very sensitive to the exact number of clusters and we get
very similar results if, for example, instead of 13 clusters we have 23 or 7
clusters. We show the effect of choosing different number of
  clusters or partitions in Appendix \ref{app}.

\subsection{Choosing the partition boundaries} 
\begin{figure}
\resizebox{\hsize}{!}{\includegraphics{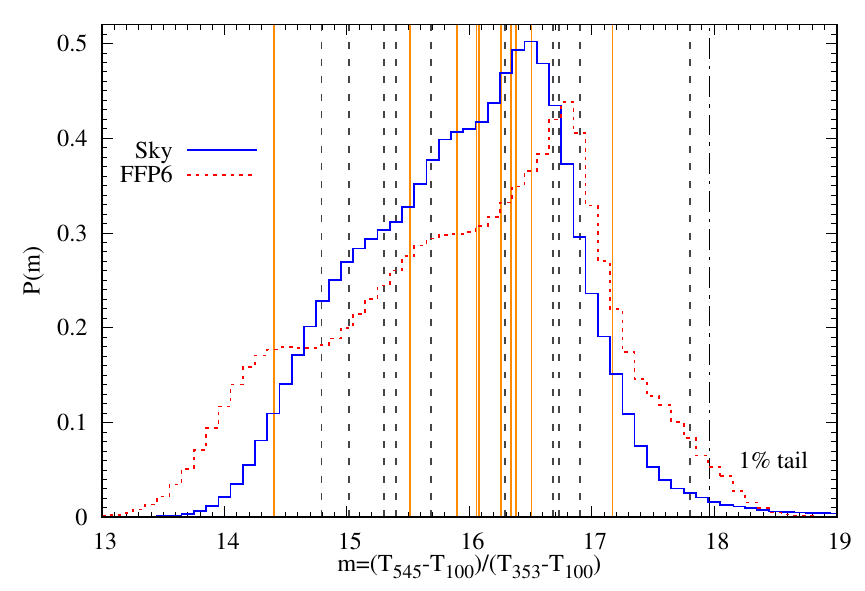}}
\caption{\label{Fig:partition}PDF of the measure $m$ for the FFP6
  simulations and Planck PR2 sky. We show two random realizations of 
  partitions by vertical solid and dashed lines. The rightmost partition (dot-dashed curve) is a fixed
  partition containing the $1\%$ of the most contaminated pixels.}
\end{figure}
We want to cluster the pixels so that the pixels closer together in $m$
value are in the same cluster. The simplest  way would be to divide the $m$
range spanned by the data into equal intervals. This will however lead to a
problem of too few pixels in  the clusters away from the main peak of the
PDF with most of the pixels in a single cluster. We instead sort the pixels
according to their $m$ value.  We can now partition the sorted pixel array so
that each cluster has equal number of pixels. However any fixed clustering
seems a bit arbitrary, since the pixels at the boundary would be closer in
their $m$ value to the pixels just across the cluster boundary than the
pixels in the same cluster near the other boundary of the cluster. 

What we 
would like is that the ILC solution at each pixel is influenced most by the
pixels nearest to it in $m$ value and less and less as we move away in
$m$. Equivalently the pixels closest to a given pixel in the $m$-sorted
pixel array should most strongly influence it and the pixels further and further away in the
array have lesser and lesser  influence in deciding the ILC solution or
weights. We therefore choose the following algorithm for the clustering: 
\begin{enumerate}
\item Randomly choose partitions uniformly between the
minimum pixel position in the sorted array $\pmin$ and the maximum
$\pmax$, the minimum and maximum corresponding to the cutoffs we placed at the tails of the PDF $P(m)$.
\item Perform ILC in each cluster.
\item Repeat the above steps $N$ times, each time with a new random
  realization of partitions. 
\item The final solution at each pixel is obtained by average of the $N$
  solutions.
\end{enumerate}
\begin{figure}
\resizebox{\hsize}{!}{\includegraphics{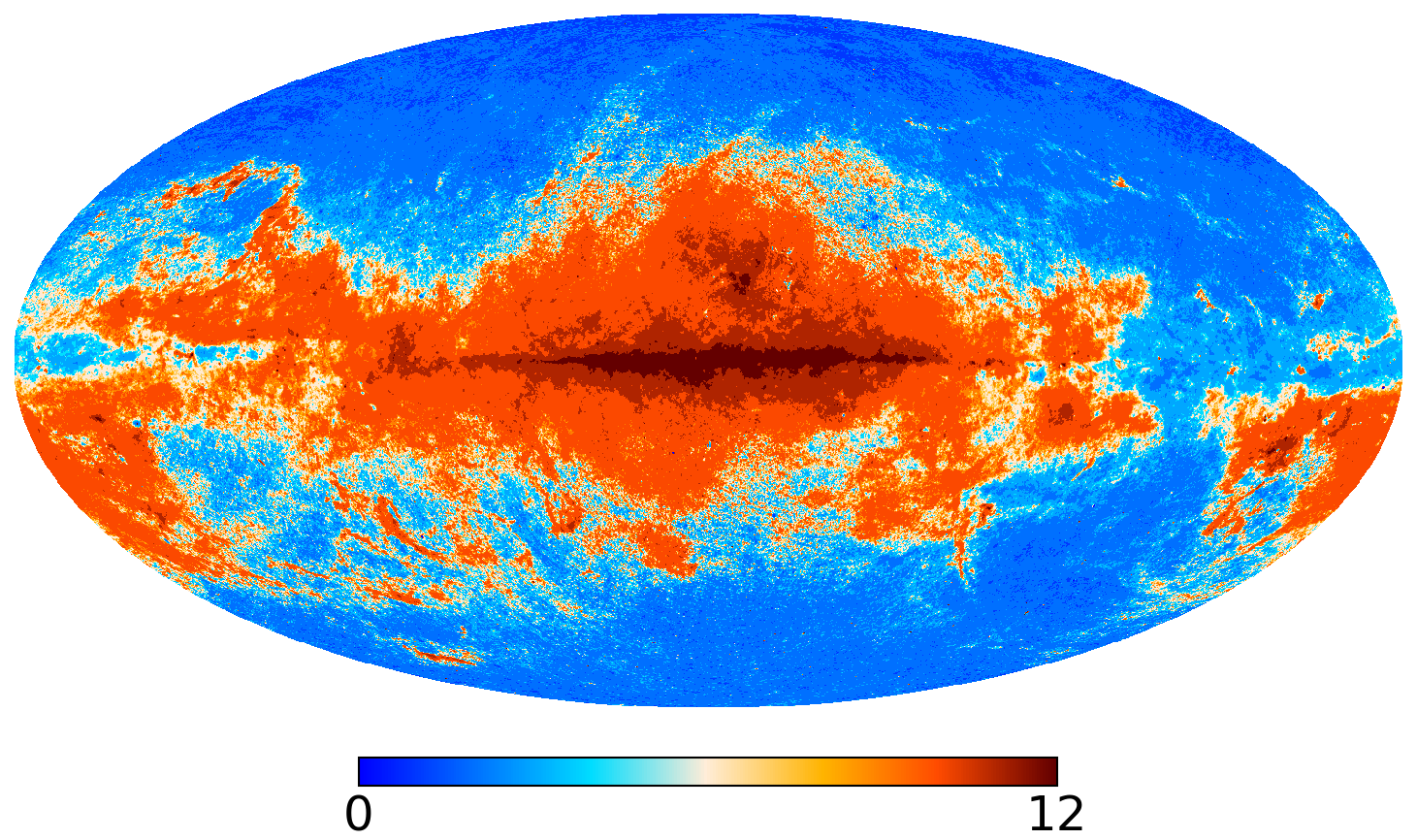}\includegraphics{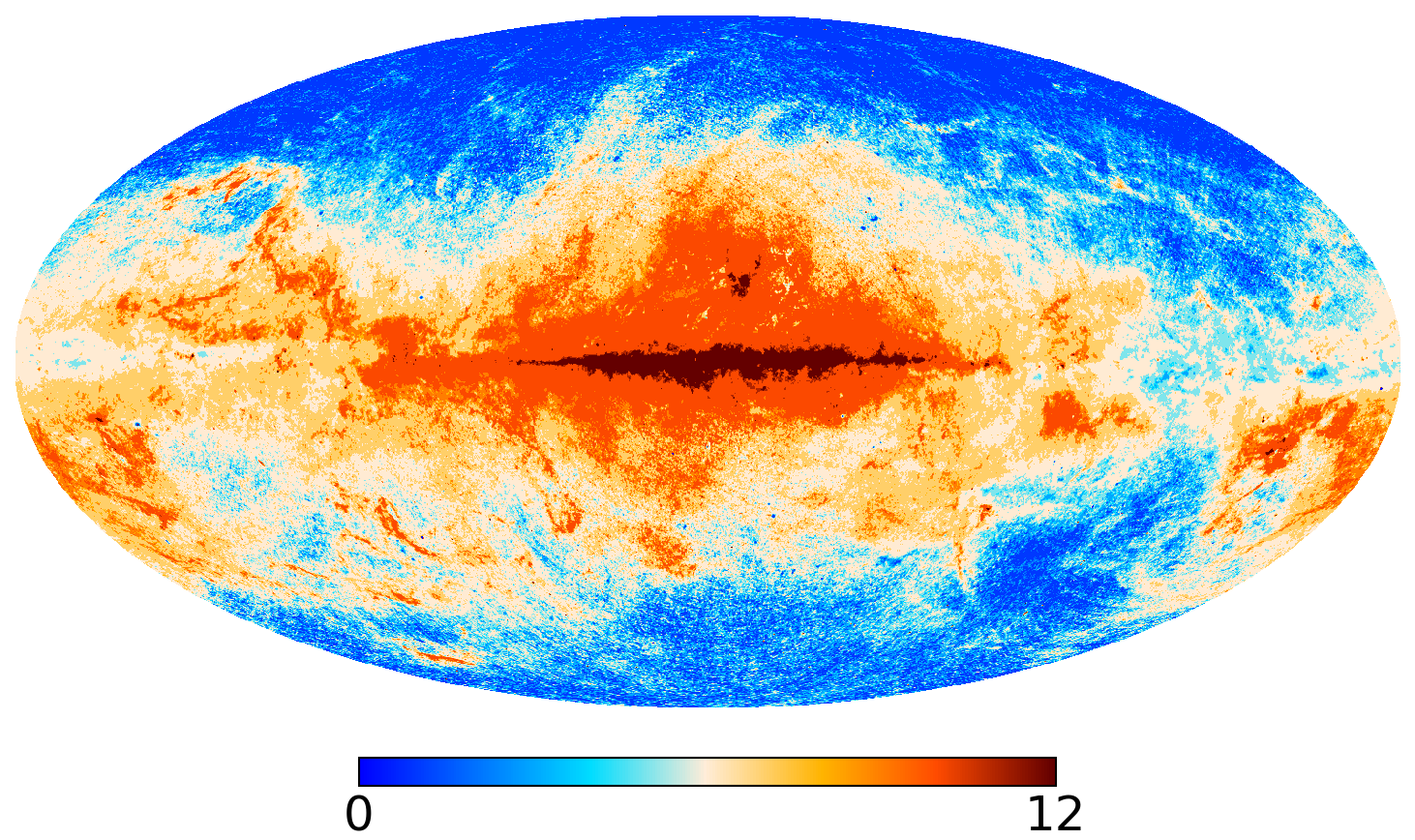}}
\caption{\label{Fig:partitionmap}Two random realizations of clustering for
  Planck PR2 sky. Each pixel value, between 0 and 12, corresponds to the cluster number it
  belongs to.}
\end{figure}

We show two such random partitionings in Fig. \ref{Fig:partition}. We see
that the some partitions are quite small while others are large.
The advantage of the above procedure is that there is automatically a larger
probability that pixels closer to each other would find themselves in
the same cluster more often compared to pixels further away which is what
we want. Also, the partition boundaries are automatically blurred as they
are randomly chosen each time and there is no need for any other
smoothing. We find that the solution converges very rapidly as we increase
$N$ and after $N=100$ there is no noticeable change in the final
solution. The convergence with the increasing $N$ is shown and further
discussed in Appendix \ref{app}. In particular the change in FC-ILC
solution when doubling $N$ from $100$ to $200$ is $< 1 ~{\rm mu K}$ at
$\Nside=512$ resolution, except
for a few extremely contaminated pixels in the two bad clusters, indicating
that the algorithm has converged to a desired level. We therefore choose this value of $N$. The high values of $m$
also correspond to the most contaminated regions on the sky near the
Galactic center which also have more complicated foreground properties. We
do not want these regions to influence the ILC solution in the neighbouring
regions. We therefore put the pixels belonging to the $1\%$ high $m$ tail of $P(m)$
in a fixed cluster (see Fig. \ref{Fig:partition}) and choose random
partition boundaries  for only the remaining $11$
pixel clusters. The maps of two random realizations of clustering are shown
in Fig. \ref{Fig:partitionmap} .

\section{Internal linear combination: FC-ILC}
We want to separate out the cosmological signal of interest, in this paper
the CMB, from other cosmological components and galactic foregrounds in
each pixel cluster or partition. We note that similar to the usual ILC, our
method can also be used to look for any other signal with known spectrum,
the generalization usually going by the name of Generalized ILC in CMB
literature. The generalization is quite trivial and just involves transforming
the data into units in which the desired signal is the  same in all frequency
channels and then applying the ILC method described below.  

The observed data, $T_{i}(p)$ in each pixel $p$ in
frequency channel $i$ can be written, in $K_{\rm CMB}$ units as 
\begin{align}
T_i(p)=\scmb(p)+ f_i(p) + n_i(p),
\end{align}
where $\scmb$ is the CMB temperature independent of frequency but a
function of pixel number $p$, $f_i$ is sum of all other components which
have a different spectrum than the CMB and are therefore a function of the
frequency channel $i$ as well as the pixel number $p$ and $n_i(p)$ is the noise in
channel $i$ and pixel $p$. We will be using HEALPix (Hierarchical Equal
Area Iso Latitude pixelation of the sphere \cite{healpix}) pixelization and $p$
denotes the pixel number in one of the HEALPix ordering schemes. We want to construct a linear combination of
data $T_i$ which preserves the desired signal, $\scmb$, i.e. we want to find weights, $\wia$ for
each pixel cluster $\alpha$ such that,
\begin{align}
\hscmb(\pa) &= \sum_i \wia T_i(\pa) \label{Eq:s}\\
\sum_i \wia & = 1\label{Eq:w},
\end{align} 
where $p^{\alpha}$ is the pixel belonging to partition $\alpha$, and
$\hscmb$ is our solution for the CMB signal, 
\begin{align}
\hscmb(\pa)=\scmb(\pa)+\sum_i\wia \left[f_i(\pa)+ n_i(\pa)\right].
\end{align}
 We want to choose $\wia$ such
that the foregrounds and noise contribution to $\hscmb$, the second term in
the above equation, is minimized. Put
another way, we want $\hscmb$ to be as close to $\scmb$ as possible. This
suggests the following cost function to be minimized,
\begin{align}
C'&=\sum_{\pa}\left[\hscmb(\pa)-\scmb(\pa)\right]^2\nonumber\\
&=\sum_{\pa}\left[\sum_i\wia \left(f_i(\pa)+n_i(\pa)\right)\right]^2\label{Eq:cp}.
\end{align}
Clearly, when the weights are chosen so that $C'$ is minimized  the
residuals from the combination of foregrounds and noise in our estimator
$\hscmb$ are minimum. However, we cannot compute the cost function $C'$
from the data since it depends upon the true signal $\scmb$. We have access to only $\hscmb$. We can therefore
try to minimize the following cost function, which can be computed from the
data, as the next best thing,
\begin{align}
C&=\sum_{\pa}\left[\hscmb(\pa)\right]^2\nonumber\\
&=\sum_{\pa}\scmb(\pa)^2 + C' + 2\sum_{\pa} \scmb(\pa) \sum_i\wia \left[f_i(\pa)+ n_i(\pa)\right].\label{Eq:C}
\end{align}
Since the first term does not depend on $\wia$, minimizing $C$ will be
equivalent to minimizing $C'$ if the last term vanishes, i.e. the signal is
uncorrelated with the foregrounds. We note that the instrumental noise in
most experiments would be uncorrelated with the signal and the foregrounds
but there maybe some chance correlation between the foregrounds and the
signal given the finite amount of data, especially on large scales where
the number of independent modes is very small. This chance correlation
means that the last term in Eq. \ref{Eq:C} is not zero and minimizing $C$
is not equivalent to minimizing $C'$. We will therefore get a biased result,
 with the ILC solution also removing the part of the signal correlated with
 the foregrounds. This is known as the ILC
bias \cite{te1996,dc2007,nilc} and is the price we pay for not knowing a priori the spectrum
of the foregrounds in such blind algorithms. The ILC bias results in a
solution which has less power than the actual signal. We find that in FFP6
simulations the ILC bias is significant, with our algorithm, only for the quadrupole ($\ell=2$)
and is negligible for all higher multipoles. However, it can start affecting
higher multipoles if we increase the number of partitions to a
very high number. 
\begin{figure}
\resizebox{\hsize}{!}{\includegraphics{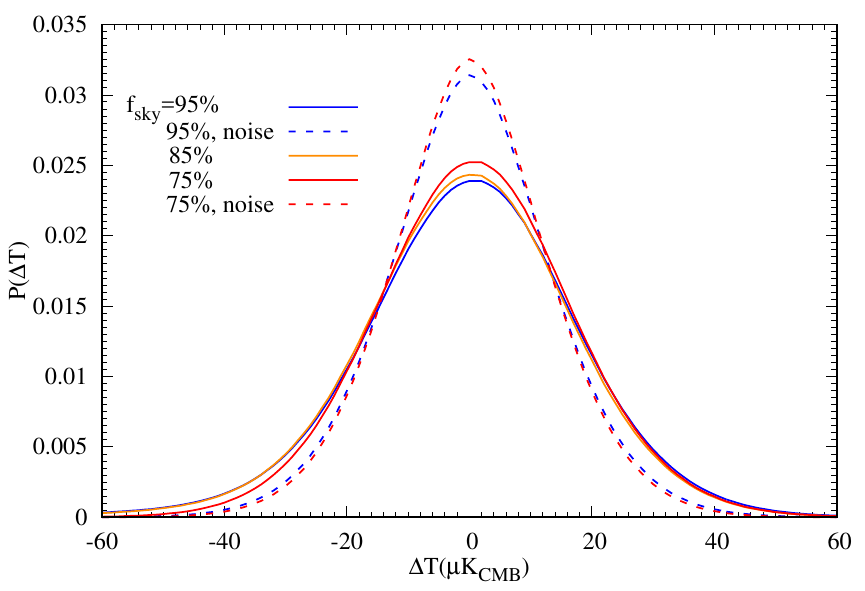}}
\caption{\label{Fig:psmpdf}The probability density function of the
  difference between the input CMB map and the ILC solution, after removing
the monopole, the dipole, and the quadrupole at full resolution $\Nside=2048$. The red and blue dashed lines are the noise PDF
from half ring half difference maps for $75\%$ and $95\%$ sky fractions
respectively at the map resolution of $5'$. For most pixels, the difference
is consistent with the noise, the small foreground contamination shows up
in the tails of the PDF.}
\end{figure}

\begin{figure}
\begin{subfigure}{0.49\hsize}
\resizebox{\hsize}{!}{\includegraphics{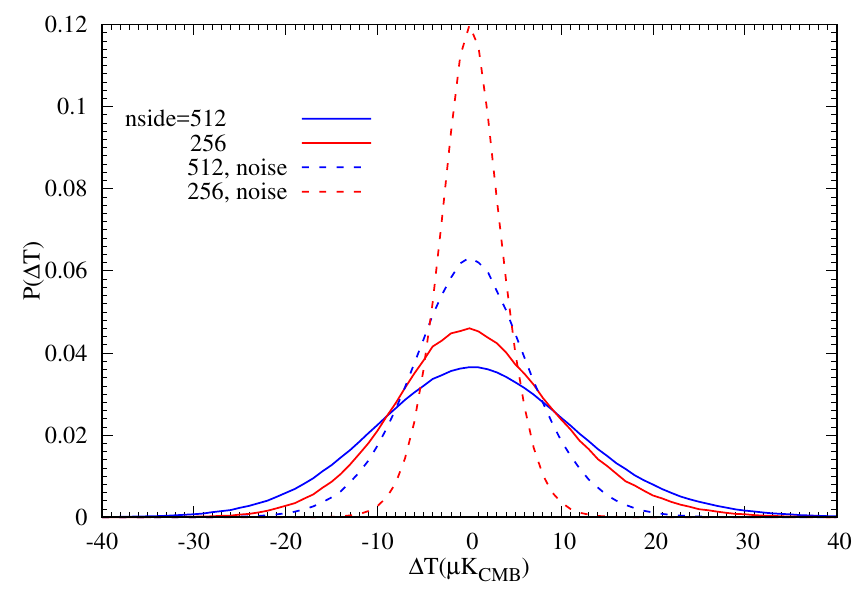}}
\caption{$75\%$ sky fraction}
\end{subfigure}
\begin{subfigure}{0.49\hsize}
\resizebox{\hsize}{!}{\includegraphics{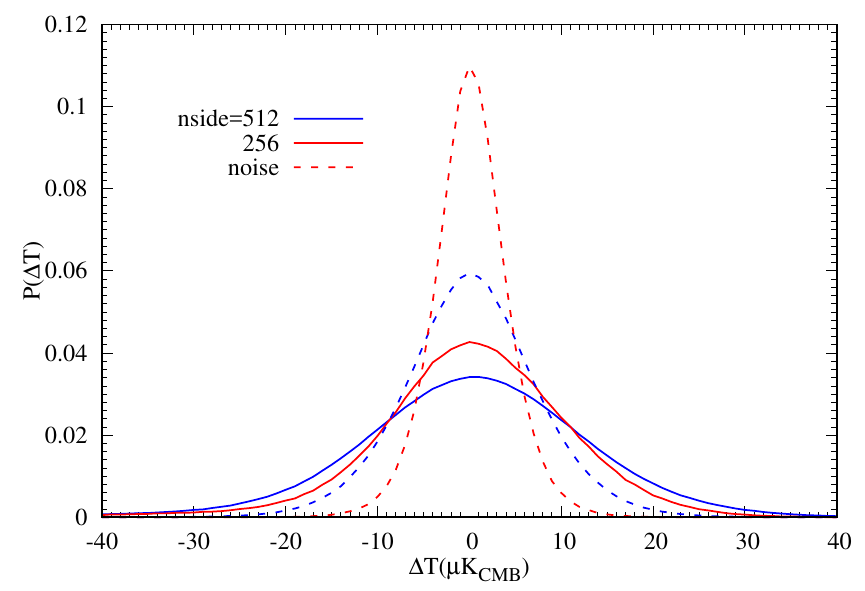}}
\caption{$95\%$ sky fraction}
\end{subfigure}
\caption{\label{Fig:psmpdf2}The probability density function of the
  difference between the input CMB map and the ILC solution, after removing
the monopole, the dipole, and the quadrupole at resolutions $\Nside=512,256$. The red and blue dashed lines are the noise PDF
from half ring half difference maps. For most pixels, the difference
is consistent with the noise, the small foreground contamination shows up
in the tails of the PDF. The suppression of noise at lower resolution means
that more pixels are dominated by contamination as we decrease the
resolution as evident by the growing difference between the noise and CMB
pdfs. We note that the average size of a pixel at $\Nside=512$ resolution is 6.87
arcmin, close to the resolution of the map and is therefore a fair
representation of the contamination in the map.}
\end{figure}

We note that our cost function $C$ is not the covariance matrix that is
usually used in the ILC. The
covariance matrix usually minimized in standard ILC is defined by 
\begin{align}
Cov=\sum_{\pa}\left[\hscmb(\pa)-\langle\hscmb(\pa)\rangle\right]^2,
\end{align}
where $\langle\hscmb(\pa)\rangle$ is the average value of $\hscmb$ of the
pixels in the partition $\alpha$. In general $C$ and $Cov$ would have
different foreground properties and will give different results. Which of
the two is better would depend on the signal as well as the nature of
data. For Planck data, we find that using $C$ as the cost function to be
optimized gives slightly better results for the CMB both in the FFP6
simulations as well as on the real sky. Since in Planck data, with the zero
level set to the Galactic zero levels, the CMB already has a zero mean
while most of the foregrounds are strictly positive, we can expect that
trying to remove all foregrounds, including the average foreground signal, in a cluster
would give better results. Another possibility is that the Covariance matrix
has slightly more complicated foregrounds properties resulting in a
slightly worse solution. Also, minimizing the
$Cov$ would leave a residual average bias in each cluster, since we are
subtracting the average signal from the partial sky, which may bias
the large scale modes. We therefore use $C$ in Eq. \ref{Eq:C} as our
cost function to be optimized.

We now define the matrix $\mathbf{D}$ as
\begin{align}
\mathbf{D^{\alpha}}_{ij}&\equiv \sum_{\pa}T_i(\pa)T_j(\pa).
\end{align}
The usual minimization using Lagrange multipliers \cite{Eriksen2004} gives
the ILC solution in each partition,
\begin{align}
w_i^{\alpha}=\frac{\sum_j\mathbf{D^{\alpha}}^{-1}_{ij}}{\sum_{ij}\mathbf{D^{\alpha}}_{ij}^{-1}}.
\end{align}

\begin{figure}
\resizebox{\hsize}{!}{\includegraphics{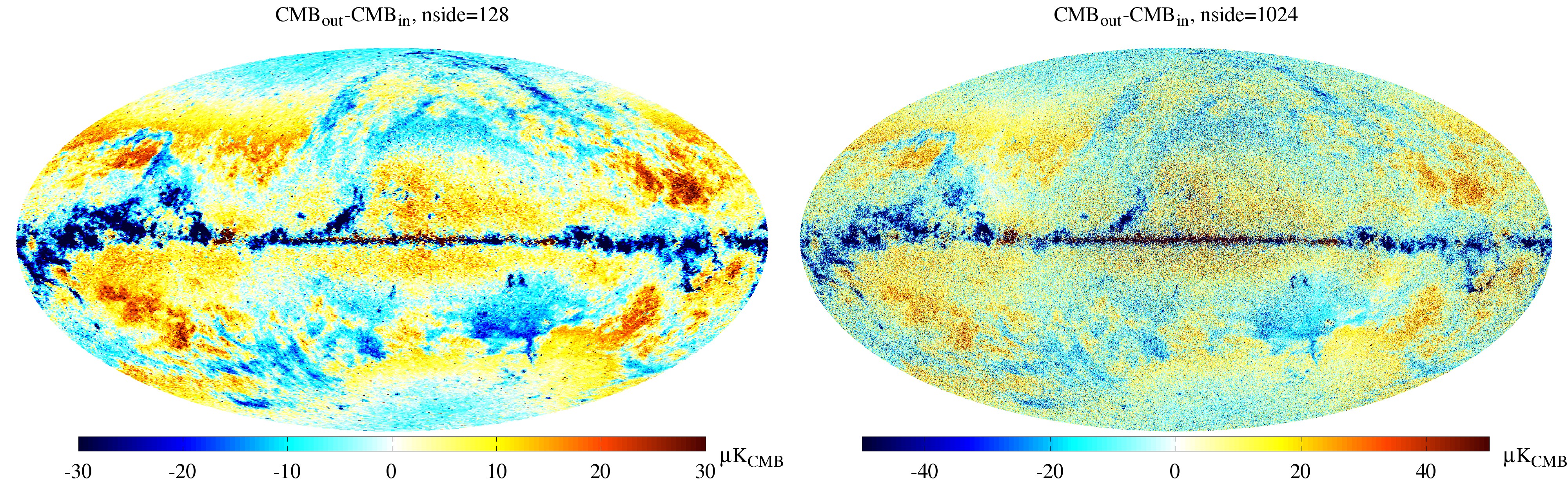}}
\caption{\label{Fig:psmdiffmaps}The difference between the input map and
  FC-ILC solution for the FFP6 simulation at HEALPix resolution
  $\Nside=128$ and $\Nside=1024$. For most of the sky the difference is less
than $10\mu$K for $\Nside=128$.}
\end{figure}
As explained earlier, we choose partitions randomly and get the ILC
solution in each pixel $N=100$ times and the final solution is the average of
$N$ solutions. 

\subsection{Combining information from maps with different resolutions}
Before we apply our algorithm, we must also find a way to optimally combine
the information from maps at different frequencies which in any CMB
experiment usually have different resolutions. This poses a
conundrum for the pixel space component separation algorithms. Noise in a
frequency channel affects the weight it gets. However, noise per pixel
depends on the resolution of the map. We should therefore first rebeam all
maps to a common resolution. If
we rebeam all maps to a common resolution corresponding to the lowest
resolution map, we loose small scale information. If we rebeam all maps to
highest resolution maps, the noise in lower resolution maps would be
boosted and the ILC solution will down weight these maps heavily,
effectively loosing any information contained in these maps. In the
harmonic space methods this is not a problem since after all maps have been
 rebeamed to the highest resolution, the noise in each map
would shoot only on scales smaller than the beam size or at high $\ell$ . Therefore, for each
$\ell$ mode, all channels with sufficient signal to noise ratio provide
information and contribute to the ILC solution. We want the same advantage
but in pixel space. To achieve this, we have come up with the following
algorithm. 
\begin{enumerate}
\item Rebeam all maps to lowest resolution map and apply FC-ILC
  algorithm. This solution will have the lowest foregrounds at that resolution
  since it uses the information from the maximum number of channels. Let's
  label the ILC solutions by index $a$, i.e. $\hscmb^a$, with $a=1$ corresponding to the
  lowest resolution, and $a>b$ implying resolution of map $a$ is higher
  than map $b$. We will denote the corresponding (Gaussian) beams with $b_{\ell}^a$.
\item Now drop the lowest resolution channel, rebeam all maps to the next
  higher resolution and obtain a new ILC solution.
\item Repeat step 2 until not enough frequency channels remain to get a
  viable ILC solution
\item Combine the $n$ solutions at different resolutions $\hscmb^a$ in
  harmonic space, i.e. $a_{\ell m}^a$, to get
  the final solution $\hscmb^f$, or in harmonic space $a_{\ell m}^f$, with the resolution corresponding to the
  highest resolution channel with beam $b_{\ell}^n$, as follows:
\begin{align}
a_{\ell m}^f= 
b_{\ell}^n\left[a_{\ell m}^1+(1-b_{\ell}^{1})\left( a_{\ell m}^{2} + (1-b_{\ell}^{2})
\left(  a_{\ell m}^{3} + .... + \left(1-b_{\ell}^{n-1}\right)\frac{a_{\ell m}^n }{b_{\ell}^n}\right)\right)\right].
\end{align}
In the last term in nested brackets in the above equation, we correct the
highest resolution solution with its beam and then use the resulting
$a_{\ell m}$ to fill in only the information not present in the next lower
resolution map, hence the factor of $(1-b_{\ell}^{n-1})$. We then use the
result to fill in only the information not present in the next lower
resolution map and so on until the lowest resolution map. The final
result  is multiplied by $b_{\ell}^n$ to get the final map at the highest
resolution possible. Note that $b_{\ell}^a$ are the beam functions of the
rebeamed maps and we  use Gaussian beams for rebeaming. The beams used for
different Planck channel combinations are given in Table \ref{Tbl:beams}.
\end{enumerate}

\begin{table}
\begin{center}
\begin{tabular}{|c|c|c|}
\hline
 Min Freq - Max Freq  &Number of channels  & Resolution (Gaussian Beam FWHM)\\
\hline
70 GHz - 545 GHz & 6 & 15' \\
100 GHz - 545 GHz & 5 & 10' \\
143 GHz -545 GHz & 4 & 7.5'\\
217 GHz - 545 GHz & 3 & 5'\\
\hline
\end{tabular}
\caption{\label{Tbl:beams}Planck frequency channel combinations and the
  rebeaming 
  resolution used for FC-ILC solutions presented in this paper. All maps
  are rebeamed to have  Gaussian beams.} 
\end{center}
\end{table}
\subsection{Masks}
We create masks with different sky fractions by first masking pixels above
a certain threshold in $m$ map to remove few $\%$ of highest $m$ pixels,
which also masks most of the point sources and then extend this mask by
removing the pixels above a threshold in the $545$GHz map after smoothing
it with a Gaussian beam of $90'$ FWHM. All masks for calculating the power
spectrum are apodized by a Gaussian replacing  $1$s in the mask by
$1-\exp\left(-9\theta^2/(2\theta_{\rm ap}^2)\right)$ for $\theta <
\theta_{\rm ap}=30'$, the apodization angle, where $\theta$ is the distance of the pixel from the
nearest masked ($0$ value) pixel.
 
\subsection{Results for FFP6 simulations}
The result of applying our FC-ILC algorithm on the Planck
FFP6 simulations is shown in Fig. \ref{Fig:psmpdf} 
where we show the
probability density function (PDF)  of the difference between the input map and
the FC-ILC map at full resolution for different sky fractions after removing the monopole, dipole and also the quadrupole
since we 
expect to have a small bias in the quadrupole. For most of the pixels, the
difference is dominated by noise. In a small fraction of pixels the
foreground residuals result in a thicker tail compared to the Gaussian
noise. The noise PDF is calculated from the half-ring half-difference
maps. We show in Fig. \ref{Fig:psmpdf2} the PDFs for lower resolutions of
$\Nside=512$ and $256$. At lower resolutions the noise is suppressed
increasing the width of the CMB pdfs compared to the noise pdfs. The
resolution of $\Nside=512$ has an average pixel size of 6.87 arcmin, close
to the resolution of our maps and is therefore a fair representation of the
noise and contamination in our maps.

\begin{figure}
\resizebox{\hsize}{!}{\includegraphics{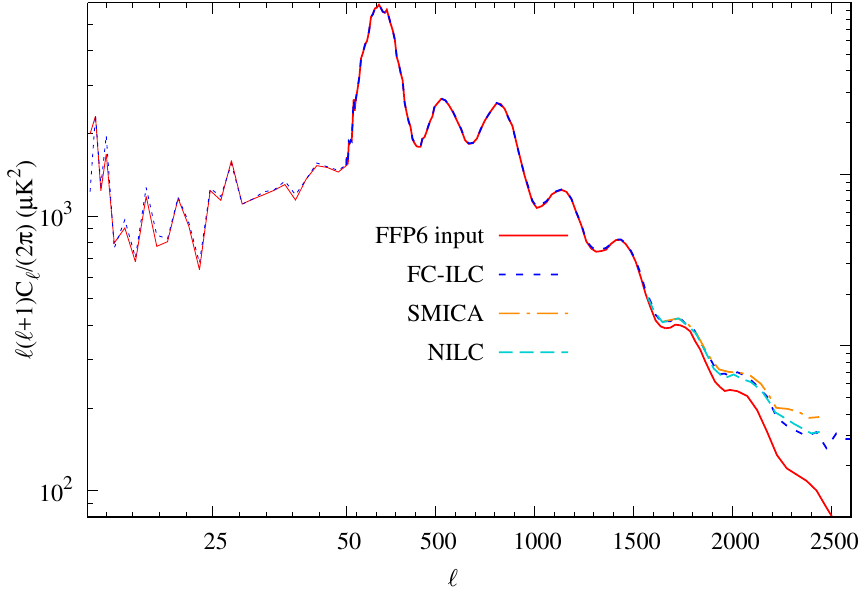}}
\caption{\label{Fig:psmcl}The FC-ILC angular power spectrum for $72\%$ sky
  fraction compared with the power spectrum of the input map with same mask
   calculated using PolSpice software \cite{ps1,ps2} which analytically deconvolves the
   mask in real space (equivalent to the harmonic space deconvolution by
   XSPECT \cite{hivon2002,xspect}). We  also show the
  SMICA and NILC solutions for FFP6 for large $\ell$ modes taken from
  \cite{planck2014}. There is ILC bias for quadrupole, which is expected,
  but it  is negligible for all higher multipoles. For highest multipoles
  our algorithm has slightly smaller residual foregrounds compared to the similar
  sky fraction for SMICA.}
\end{figure}

We show in Fig. \ref{Fig:psmdiffmaps} the difference maps at  HEALPix
$\Nside=128$ and $1024$
 and in Fig. \ref{Fig:psmcl} the angular power spectrum $C_{\ell}$ as a
 function of the multipole $\ell$. There is significant ILC bias only for
 the quadrupole. At $\ell \gtrsim 1500$ the excess is due to the residual
 foregrounds. We have plotted the SMICA $C_{\ell}$ also for the similar sky
 fraction from Fig. E.4 in \cite{planck2014}. We see that $\ell\gtrsim 2000$ FC-ILC
 slightly outperforms SMICA but performs as well as NILC. However, we should keep in mind that our masks,
 even though they cover a similar sky fraction, are different from those
 used in \cite{planck2014} and could
 be the cause of the slightly larger excess in the SMICA curve. We
 note that SMICA has the smallest contamination among all algorithms in
 Fig. 10 in \cite{planck2014} when applied to real data.

We conclude that our algorithm FC-ILC, on the FFP6 simulations, performs at
least as well as the existing algorithms, perhaps even slightly better.

\section{A new  foreground cleaned Planck CMB map}
\begin{figure}
\resizebox{\hsize}{!}{\includegraphics{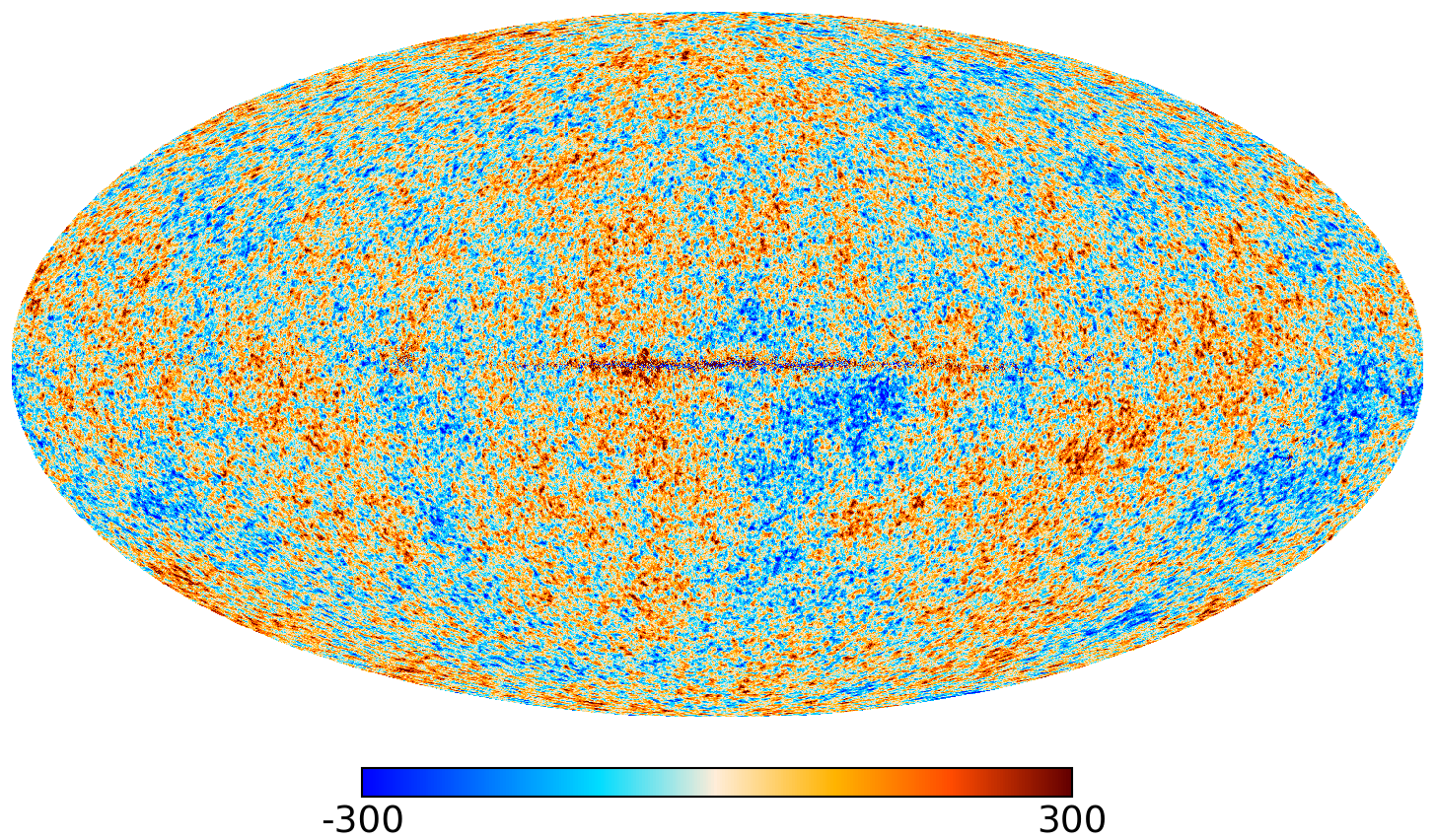}}
\caption{\label{Fig:planckmap}The Planck full mission FC-ILC map. The
  visibly contaminated region in the Galactic plane is included in the
  \emph{bad partition}, the partition number 0 in Fig. \ref{Fig:partitionmap},  consisting  $1\%$ of the most contaminated pixels.}
\end{figure}

We have applied our algorithm to the Planck PR2 release maps. The CMB map
at $5'$ resolution is shown in Fig. \ref{Fig:planckmap}. We calculate the
angular power spectrum as a cross spectrum between the half-ring maps  using the PolSpice software \cite{ps1,ps2} get the
\emph{pseudo} $C_{\ell}$ which have been 
deconvolved from the effects of the masks and corrected for the beam. PolSpice also calculates the
analytical covariance
matrix \cite[see also][]{hivon2002,xspect} which we use to estimate the error bars. We then bin the angular power spectrum  $D_{\ell}\equiv
\ell(\ell+1)C_{\ell}/(2\pi)$ as well as the covariance matrix in $\ell$ bins which increase in width with
$\ell$ for $\ell \ge 30$. 

\subsection{Angular power spectrum at $2\le \ell \le 29$}
\begin{figure}
\resizebox{\hsize}{!}{\includegraphics{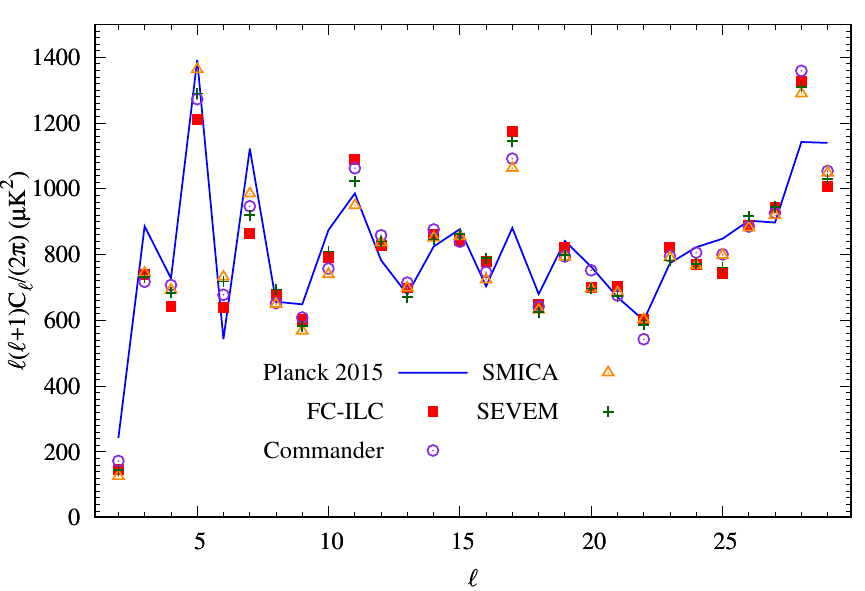}}
\caption{\label{Fig:lowl}The low $\ell$ angular power spectrum of the CMB
  calculated on a sky fraction of $95\%$ with a $30'$ apodized mask with
the publicly available   PolSpice software \cite{ps1,ps2}. The $C_{\ell}$s
have been corrected for the effect of the mask and the beam.}
\end{figure}

We show the power
spectrum for $2\le \ell \le 29$ for a sky fraction of $95\%$ in
Fig. \ref{Fig:lowl}. Also shown is the best estimate of the power spectrum
in the PR2 release of Planck collaboration \cite{planck} by solid line and
the power spectra estimated from the SMICA, NILC, Commander, and SEVEM maps
released by the Planck collaboration \cite{planck2016} and calculated with the same mask. We note that the
power spectra estimated from the cleaned CMB maps are the pseudo
$C_{\ell}$s and would therefore show some oscillations because of the
de-convolution of the mask. However, we expect these systematics due to the
incomplete sky to
be not significant since we are use a very large fraction of the sky. All maps seem to be in rough
agreement with the released power spectrum. FC-ILC  shows a slightly larger
deviation compared to the other maps. However, this deviation is for the
$\ell$ values for which the power spectrum deviates considerably  for all of the
maps shown. We note
that the ILC bias at $\ell=2$ is less compared to what we saw in the FFP6
simulations. This may mean that either the best estimate of the quadrupole
in Planck 2015 release is biased and therefore closer to the ILC solution or that the foregrounds have smaller
correlation with the CMB in the real sky compared to the simulations. Given
the consistency of Planck with WMAP \cite{wmap}, the latter is most likely the
correct explanation.

\subsection{Angular power spectrum for $30\le \ell \le 800$}
\begin{figure}
\resizebox{\hsize}{!}{\includegraphics{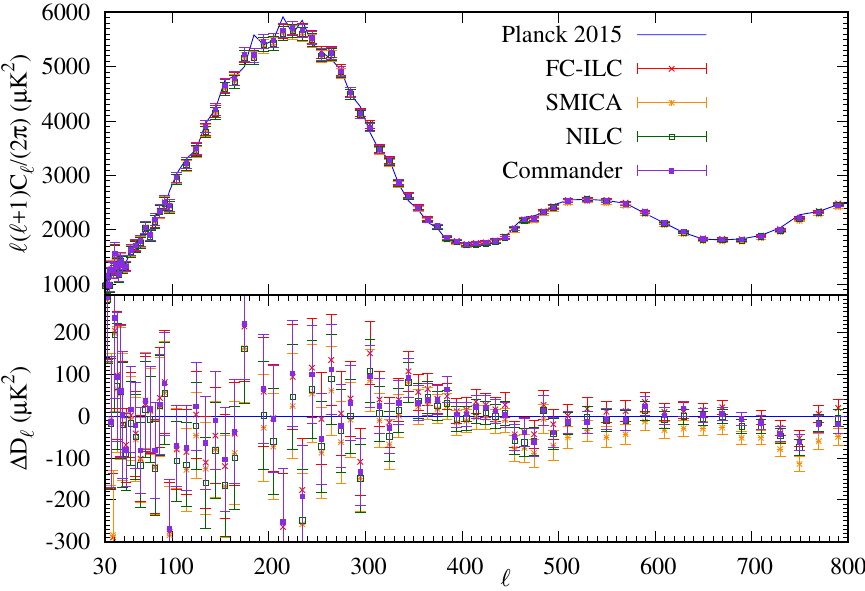}\includegraphics{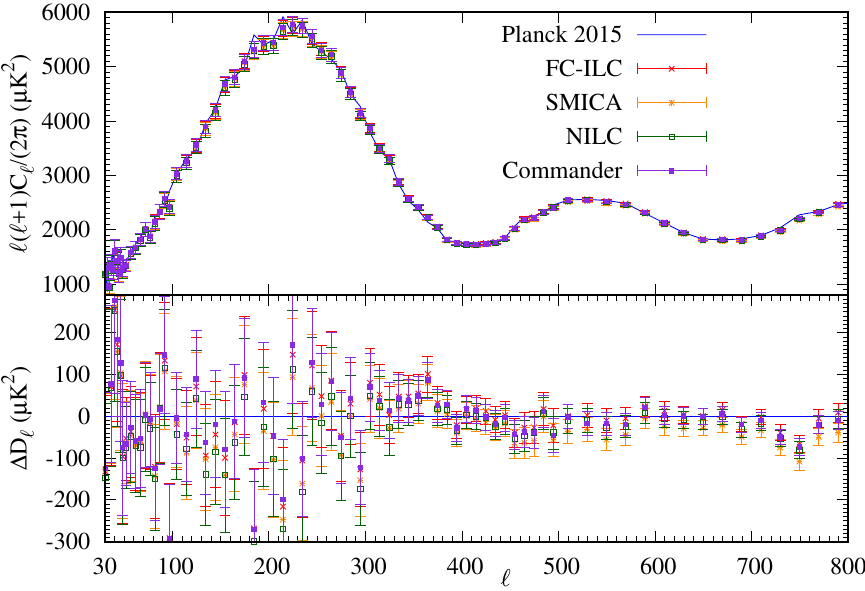}}
\caption{\label{Fig:midl1}The mid-$\ell$ angular power spectrum of the CMB
  calculated on  sky fractions of $95\%$ (left panel) and $85\%$ (right
  panel) with the masks apodized by a Gaussian with an apodization angle of  $30'$.  The bottom
panels show the difference between the $D_{\ell}\equiv \ell(\ell+1)C_{\ell}/(2\pi)$ calculated from the maps
and the Planck PR2 release $D_{\ell}$.}
\end{figure}

\begin{figure}
\resizebox{\hsize}{!}{\includegraphics{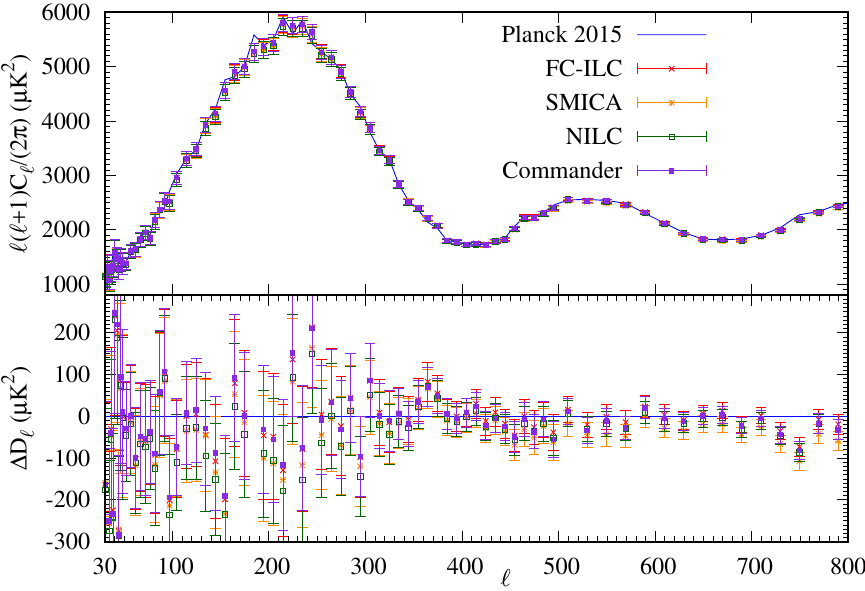}\includegraphics{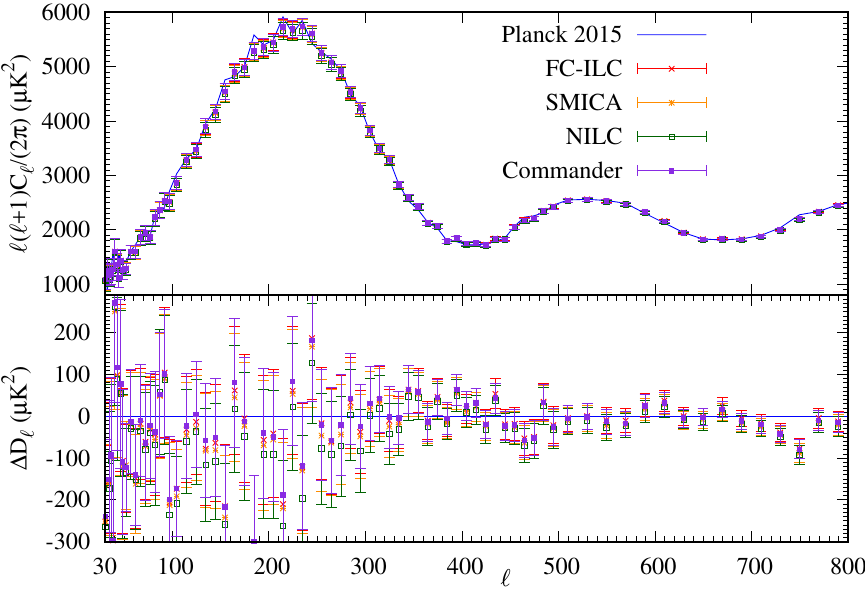}}
\caption{\label{Fig:midl2}The mid-$\ell$ angular power spectrum of the CMB
  calculated on a sky fractions of $75\%$ (left panel) and $78\%$ (Planck
  UT78  mask, right panel).  The bottom
panels show the difference between the $D_{\ell}\equiv \ell(\ell+1)C_{\ell}/(2\pi)$ calculated from the maps
and the Planck PR2 release $D_{\ell}$.}
\end{figure}

We show the binned
angular power spectrum for $30 \le \ell \le 800$ in Figs. \ref{Fig:midl1} and
\ref{Fig:midl2} for the sky fraction of $95\%, 85\%$ and $75\%,78\%$ respectively for FC-ILC
as well as Planck Commander, NILC, and SMICA maps calculated in an identical way. The $78\%$
mask is the UT78 mask released by the Planck collaboration and is a union of
masks used for different component separation methods. The solid
line is the high $\ell$ power spectrum released by the Planck collaboration
estimated from the $66\%, 57\%$ and $47\%$ cleanest portion of the sky of 100 GHz, 143 GHz and 217 GHz
maps respectively \cite{planck}. This power spectrum can therefore be taken
as approximately the \emph{true} $C_{\ell}$s of our sky. Therefore, apart
from the minor statistical fluctuations coming from having a large portion
of the sky, the $C_{\ell}$s estimated from the foreground cleaned maps
should be close to these. In particular the deviation, both positive and
negative, from the \emph{true} power spectrum can be used as a measure of
the foreground contamination in the CMB maps. 
The  SMICA  map shows significant negative bias at $ \ell\ge 500$. The bias
is much smaller in the NILC maps which lie between the Commander and SMICA
solutions for most multipoles. The  FC-ILC
and Commander maps do not show this bias  and are closer to the
\emph{clean} power spectrum released by Planck. The bias in SMICA maps
decreases for $75\%$ sky and also when using UT78 mask released by Planck as
seen in Fig. \ref{Fig:midl2}. In particular there is a constant
\emph{negative} offset
between SMICA and other maps at $\ell \gtrsim 500$.

\subsection{Angular power spectrum at $800 \le \ell \le 2500$}
\begin{figure}
\resizebox{\hsize}{!}{\includegraphics{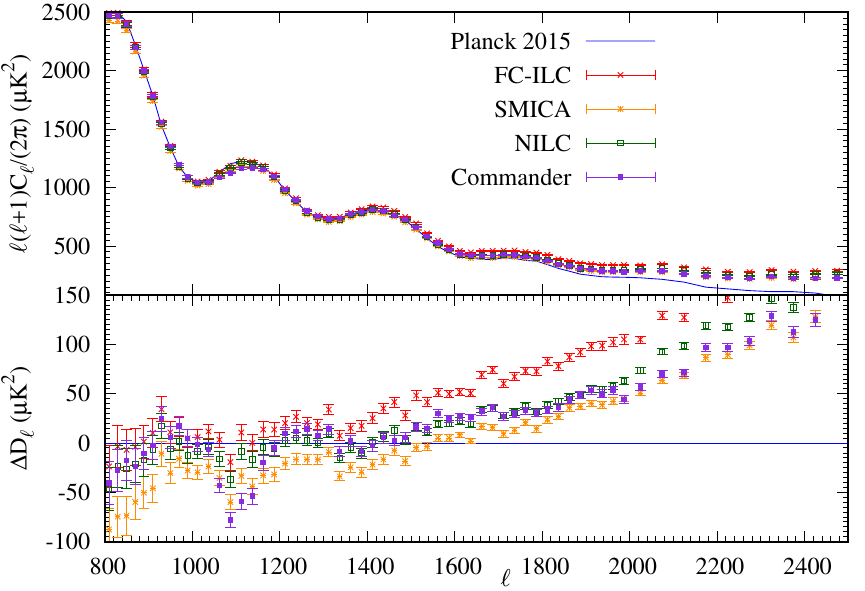}\includegraphics{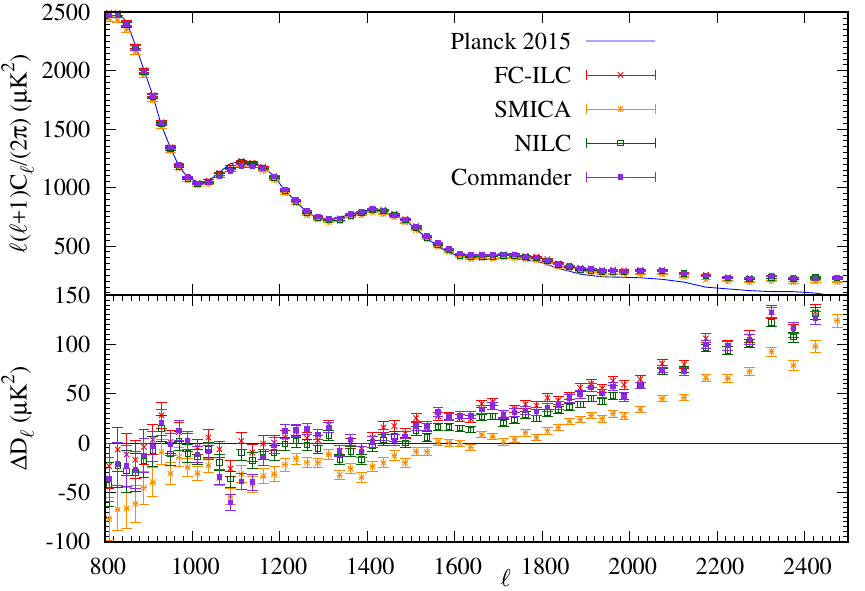}}
\caption{\label{Fig:highl1}The high-$\ell$ angular power spectrum of the CMB
  calculated on a sky fractions of $90\%$ (left panel) and $85\%$ (right panel). The bottom
panels show the difference between the $D_{\ell}\equiv \ell(\ell+1)C_{\ell}/(2\pi)$ calculated from the maps
and the Planck PR2 release $D_{\ell}$.}
\end{figure}

\begin{figure}
\resizebox{\hsize}{!}{\includegraphics{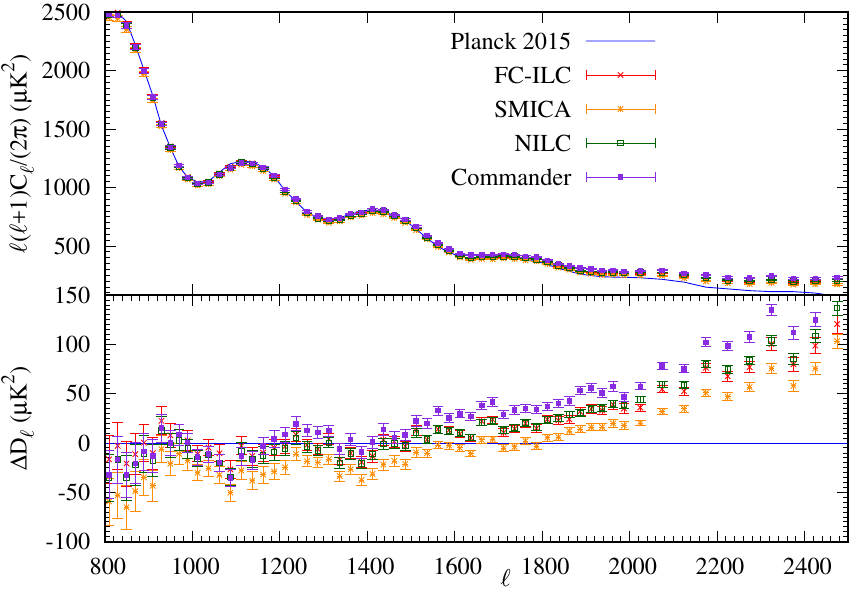}\includegraphics{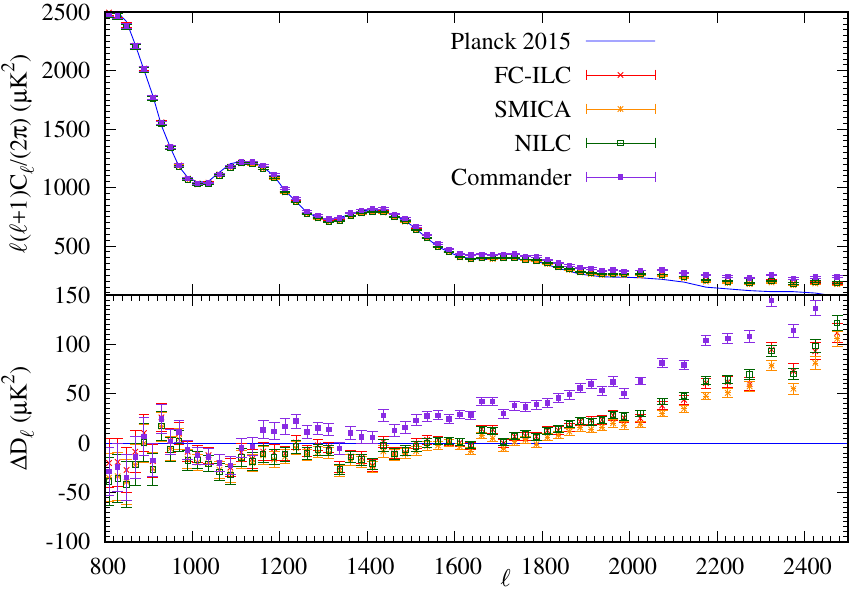}}
\caption{\label{Fig:highl2}The high-$\ell$ angular power spectrum of the CMB
  calculated on a sky fractions of $75\%$ (left panel) and $78\%$ (Planck
  UT78  mask, right panel). The bottom
panels show the difference between the $D_{\ell}\equiv \ell(\ell+1)C_{\ell}/(2\pi)$ calculated from the maps
and the Planck PR2 release $D_{\ell}$.}
\end{figure}

We show the high $\ell$ angular power spectra, for $\ell \ge 800$ in
Fig. \ref{Fig:highl1} for $90\%$ and $85\%$ sky fraction and in
Fig. \ref{Fig:highl2} for $75\%$ sky fraction and Planck UT78 mask. There
is a large negative contamination in both SMICA and Commander maps for $\ge
85\%$ sky fractions around $\ell = 1100$. Even though, looking at $\ell >
1400$ it may seem that Commander and SMICA are cleaner than FC-ILC, this is
in fact because there is a net negative bias at high $\ell$. For the $85\%$
power spectrum FC-ILC has similar excess at $\ell>1400$ as Commander but
FC-ILC does not have the negative contamination at $\ell<1400$. There is in
fact almost a constant offset between the SMICA and FC-ILC maps.
 For the UT78 mask, FC-ILC power spectrum is almost identical to the SMICA
 and NILC 
 power spectrum while Commander shows a larger excess. The NILC power
 spectrum lies between Commander and SMICA at most multipoles and is
 closest to FC-ILC solution among the Planck public CMB  maps.

\section{Conclusions and remarks about possible improvements}
We have presented a new approach to foreground cleaning and component
separation in the multifrequency CMB experiments. Our approach is to
cluster together the data into groups or partitions, such that in each
partition the foreground properties of the data are similar. This makes the
problem of component separation more tractable since in each partition the
data can be \emph{accurately} described by a small number of components,
ideally smaller than the number of frequency channels available. Our
approach differs from the existing algorithms, which also try to cluster
the data in pixel and/or harmonic space. Instead of a pre-determined
clustering of data, we estimate the foreground properties from the data
first summarizing it into a measure $m$ and then use this measure $m$ to
partition the data \emph{probabilistically}. This step of first estimating the foreground
properties, other than the amplitude, is the new ingredient. We have
implemented our approach into a algorithm called FC-ILC. At present FC-ILC
uses a single measure, constructed from two foreground dominated Planck HFI
(high frequency instrument) channels and one clean channel, and performs
the component separation in pixel
space. However, FC-ILC approach can be extended to use more than
one measure and use  harmonic space. 

The second important new ingredient of
our algorithm is to cluster the data randomly many times and use the
average solution over these many random realizations of partitions. This
probabilistic approach to the clustering of data allows solution in a
particular pixel to be influenced by all pixels close to it in terms of the
measure $m$. There are no sharp boundaries in our solutions (CMB maps)
that would  need to be artificially smoothed.

The third ingredient is that we construct many different solutions, each  with different
number of channels, and each solution at the best resolution allowed by the
channels being used. These different resolution solutions are than combined
in such a way so as to use the information from the best solutions available at a
given scale $\ell$.

We have tested our algorithm on FFP6 simulations and shown that it works as
well as can be expected, especially compared with the results of other
algorithms on the same simulations. 

We have also applied our algorithm to the Planck PR2 temperature maps and
produced new cleaned CMB maps (half-ring maps and full mission map) which
are made publicly available for use by the cosmology
community\footnote{\url{http://theory.tifr.res.in/~khatri/CMB}}. We compare
the properties of our maps with the Planck power spectrum results from the
PR2 release as well as the foreground cleaned maps made available by the
Planck collaboration. We find that while using the UT78 mask, giving
approximately $78\%$ of sky fraction for calculating the power spectrum,
our results are consistent with SMICA, NILC  and Commander maps, having similar
residual contamination as SMICA, which is slightly better than Commander at
high $\ell$. However when using a larger sky fraction, using our masks
constructed by  thresholding our measure $m$ and the 545 GHz channel map,
both Commander and SMICA show a \emph{negative} contamination. In
particular there seems to be a constant offset between the SMICA and FC-ILC
power spectra, indicating that SMICA map has been \emph{over-cleaned}. It is
overall difficult to say which of the maps is best or most free of
contamination. We hope that our independently produced maps with a
different algorithm, with similar in amplitude
if not smaller but qualitatively different residuals as the maps released by the Planck collaboration,
would be useful in testing for the effect of residual foregrounds on
cosmological parameters, in particular when studying the marginally
significant anomalies.

We have  chosen a particular clustering  method based on a very
simple one-dimensional measure of foreground shape. The shape of the
foregrounds is of course much more complicated to be captured adequately by a single
measure. We cannot therefore claim that our clustering is optimal. 
We have however shown that our simple foreground clustering approach is
reasonable in the sense that it
works as well as any of the other component separation algorithms used by
the Planck collaboration. There is however definite room and clear
direction for further
improvement and optimization, especially using machine learning to cluster
data using more than one measure,  which we defer to a future
publication. 

\section*{Acknowledgements}
I acknowledge interesting discussion with Ben Wandelt on CMB  component
separation methods. This work was supported by grant no. ECR/2015/000078 of
Science and Engineering
Research Board (SERB) of Department of Science and Technology (DST),
 Govt. of India.   This work was also supported by MPG-DST partner group
 between Max Planck Institute for Astrophysics, Garching, Germany and Tata Institute of
Fundamental Research, Mumbai, India funded by Max-Planck-Gesellschaft
(MPG). This work made use of the HEALPix software \cite{healpix} and its
python version healpy.    

\bibliographystyle{unsrtads}
\bibliography{fcilc}

\appendix
\section{Convergence of FC-ILC solution}\label{app}
\begin{figure}
\centering
\begin{subfigure}{0.49\hsize}
\resizebox{\hsize}{!}{\includegraphics{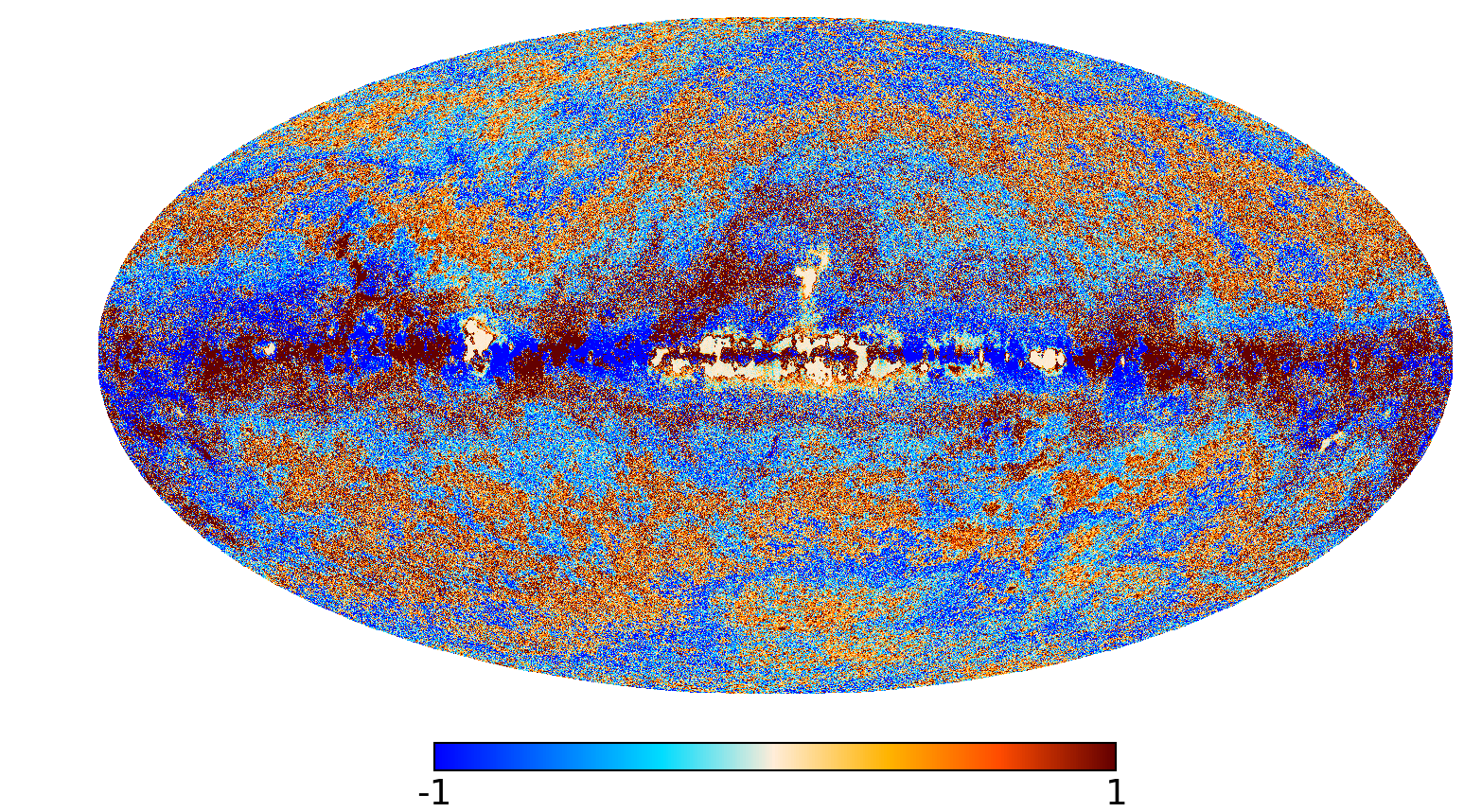}}
\caption{$(N=10)-(N=20)$}
\end{subfigure}
\begin{subfigure}{0.49\hsize}
\resizebox{\hsize}{!}{\includegraphics{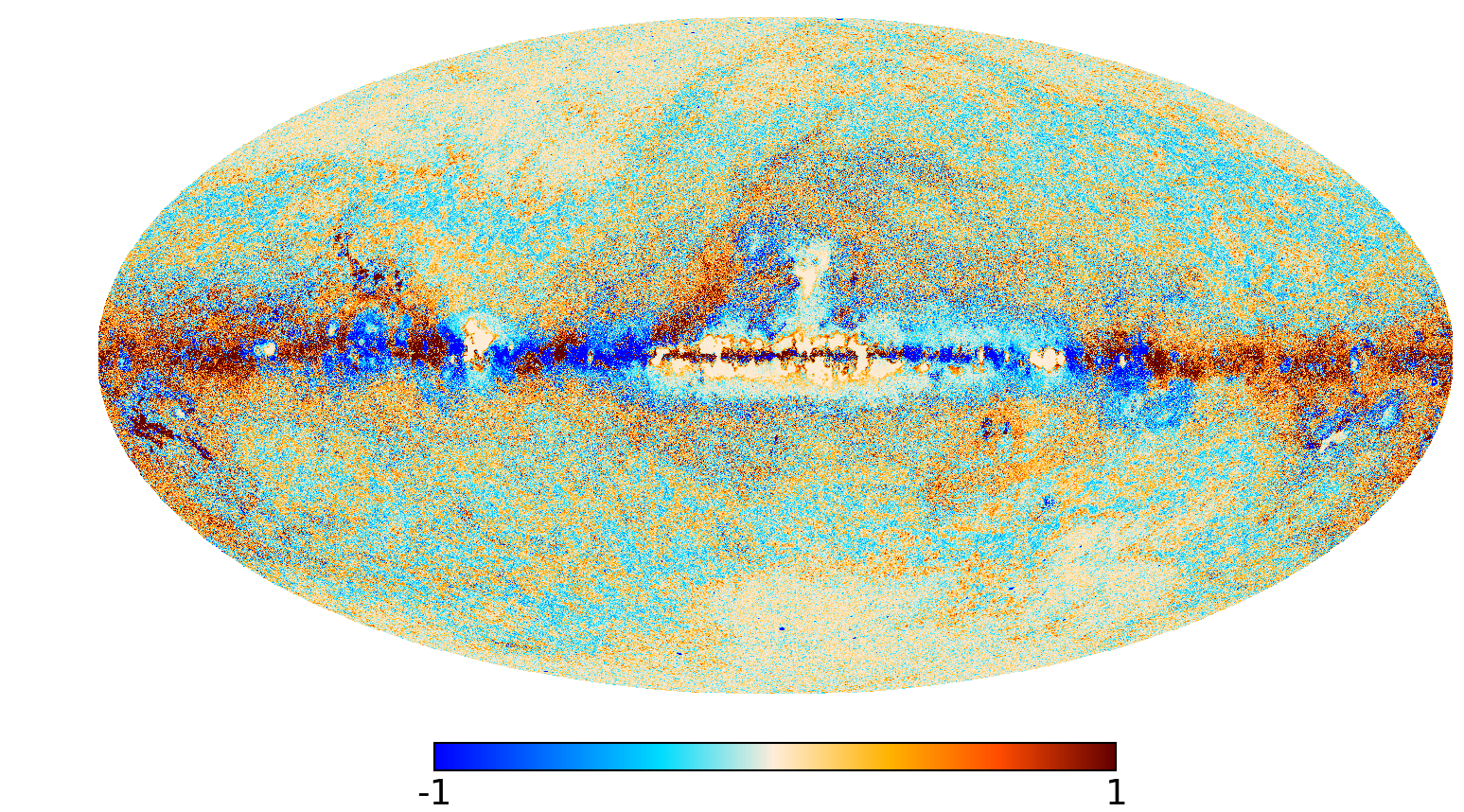}}
\caption{$(N=100)-(N=200)$}
\end{subfigure}\\
\caption{\label{Fig:n10}The change in FC-ILC solution when doubling the
  number of random realization of partitions from $N=10$ to $N=20$ (left panel) and
from $N=100$ to $N=200$ (right panel). The results are shown for the FFP6
simulation. Similar results are obtained also for the real sky data. The
difference map has been degraded from the original $\Nside=2048$ to
$\Nside=512$ for this figure. The color scale ranges from $-1~{\rm \mu K}$ to
$1~{\rm \mu K}$. Note that most of the Galactic plane  also undergoes  $<1
{\rm \mu K}$ change when going from 100 to 200 random realizations.}
\end{figure}

There are two important parameters to be chosen in the FC-ILC algorithm:
The number of random realization of partitions to be done and the number of
partitions to be chosen. The number of random realizations is a simpler choice to
make, more number of realizations would result in smoother
boundaries. Once the solution is converged, additional realizations would
not change the solution. We show in Fig. \ref{Fig:n10} the change in FC-ILC
solution when doubling the number of random realizations from $N=10$ to $N=20$ and
also the change in the solution when doubling the number of
realizations from $N=100$ to $N=200$ for the FFP6 simulation. We see that after 100 realizations, the
change in the solution in any pixel at HEALPix $\Nside=512$ is $\ll 1~{\rm
  \mu K}$ over most of the sky. We therefore fix the number of realizations
to 100. The size of a pixel at $\Nside=512$ is 6.87 arcmin, close to the 5
arcmin resolution of the CMB map and is therefore a good resolution choice
to test the convergence of the algorithm.

\begin{figure}
\centering
\begin{subfigure}{0.49\hsize}
\resizebox{\hsize}{!}{\includegraphics{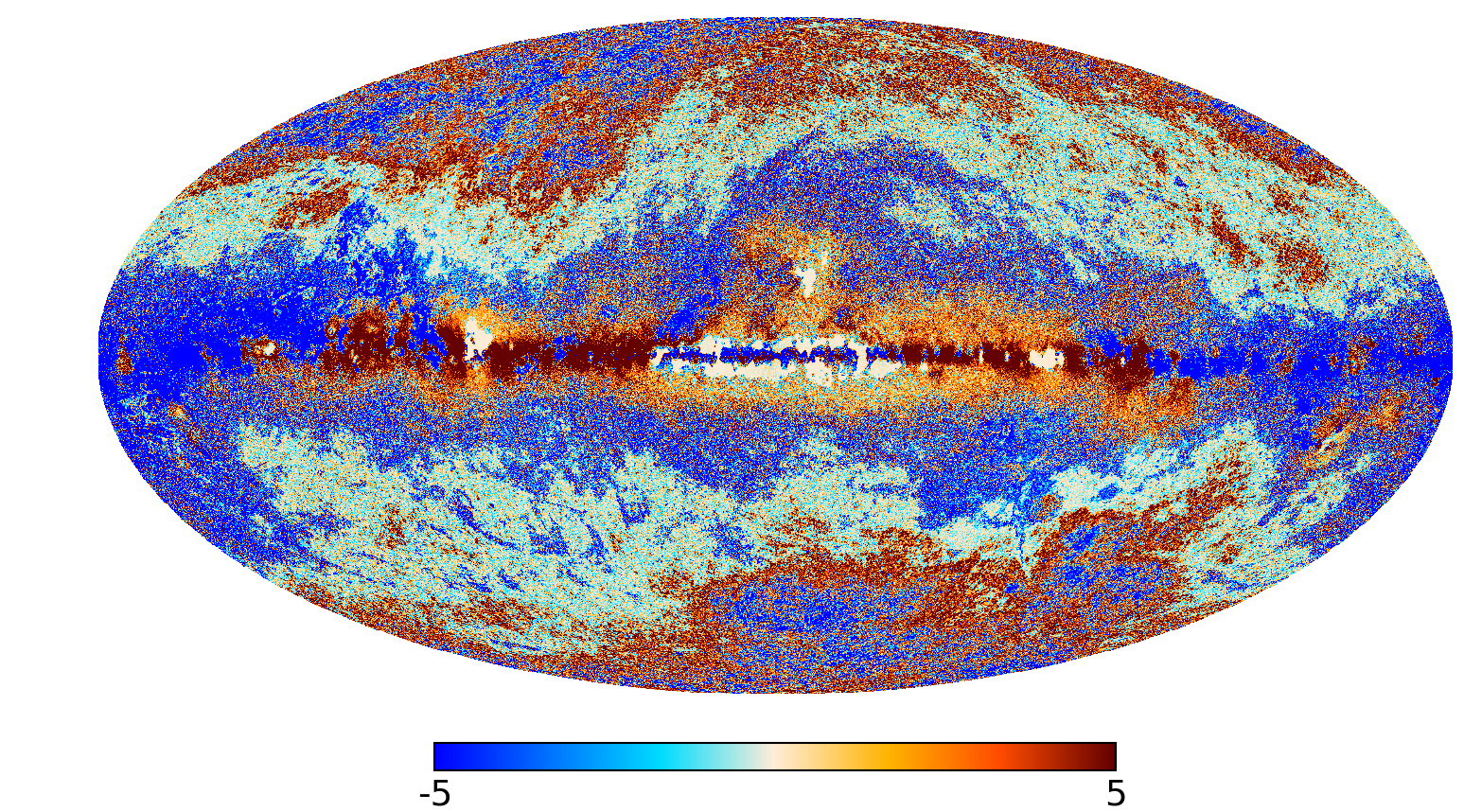}}
\caption{5 clusters}
\end{subfigure}
\begin{subfigure}{0.49\hsize}
\resizebox{\hsize}{!}{\includegraphics{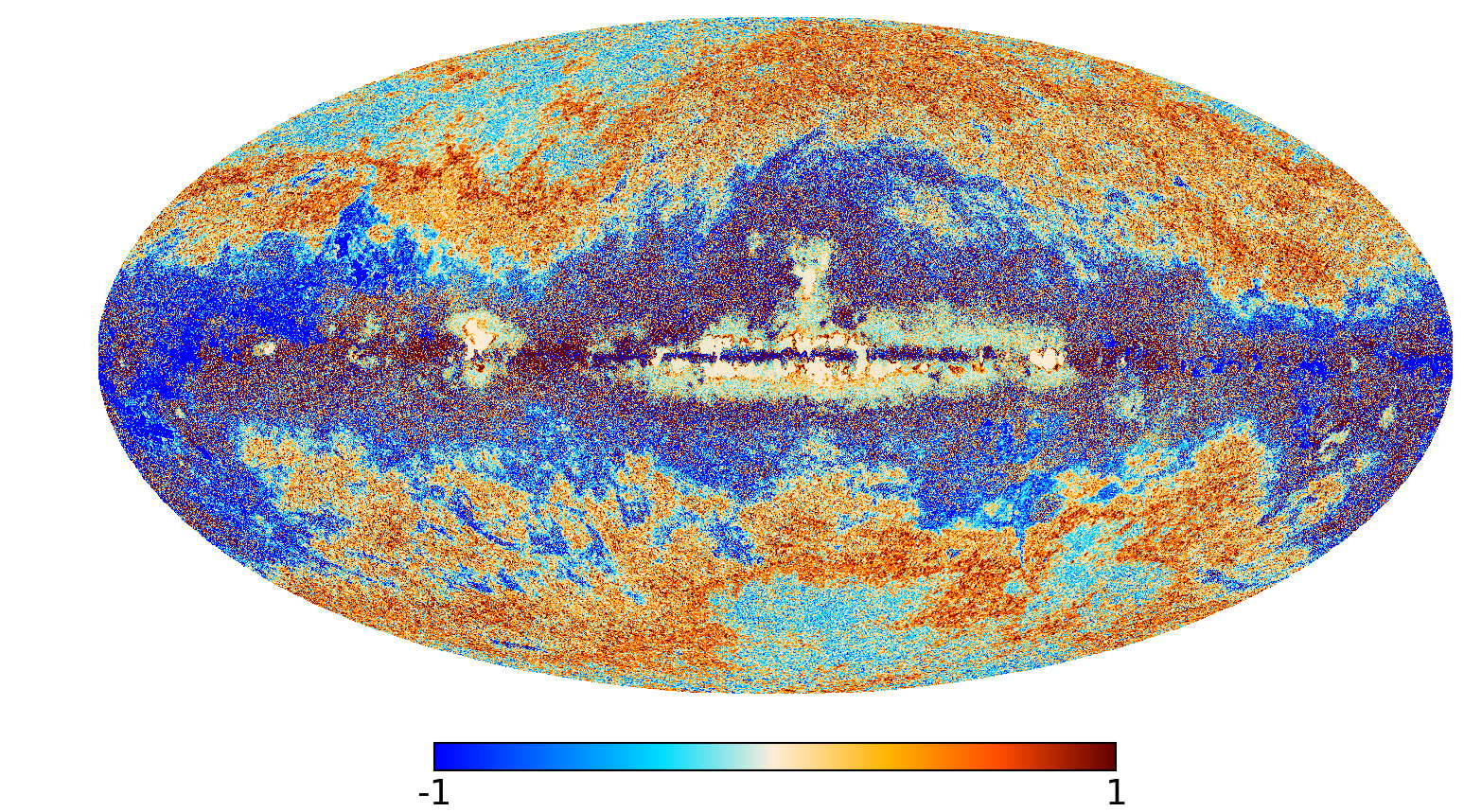}}
\caption{11 clusters}
\end{subfigure}\\
\begin{subfigure}{0.49\hsize}
\resizebox{\hsize}{!}{\includegraphics{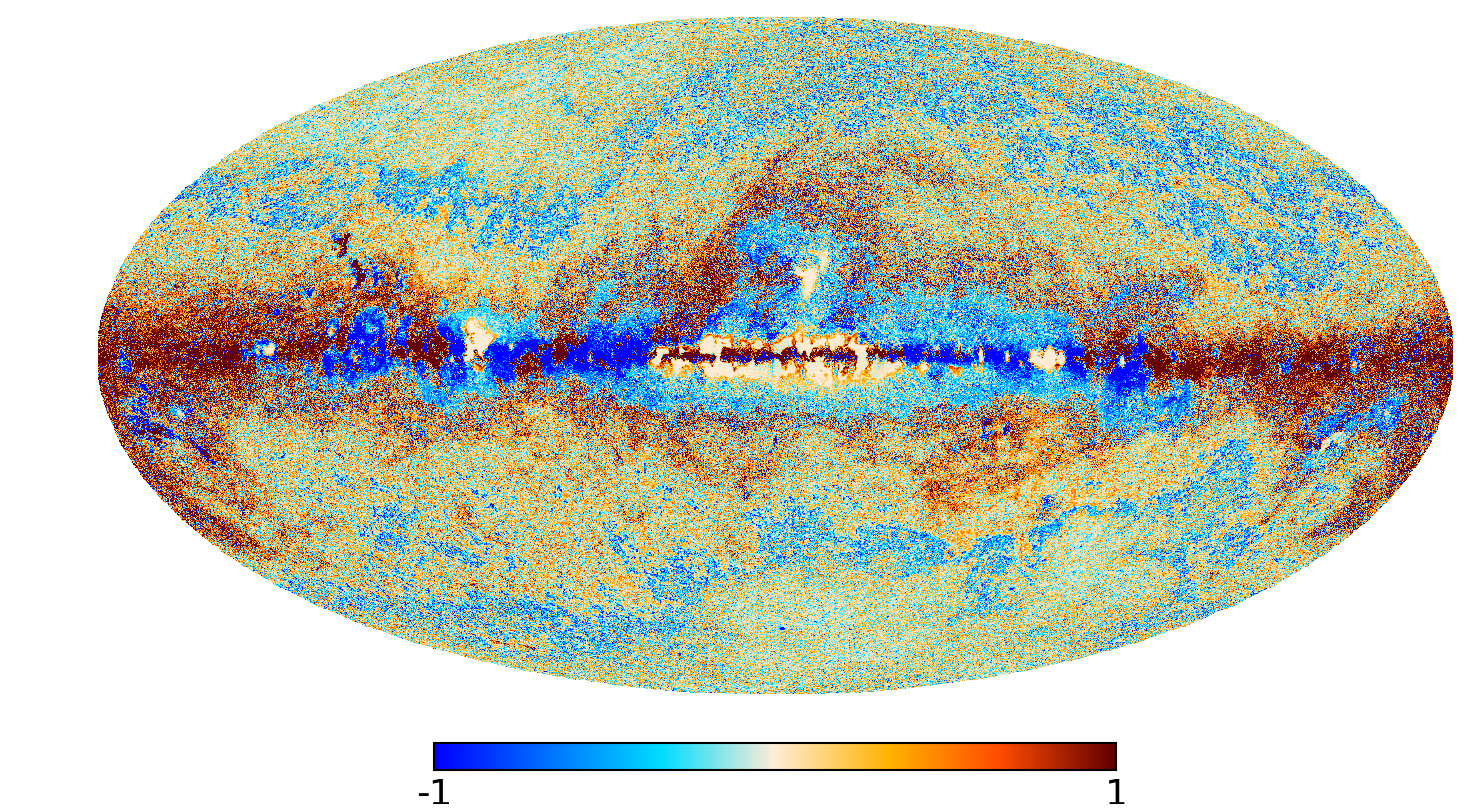}}
\caption{15 clusters}
\end{subfigure}
\begin{subfigure}{0.49\hsize}
\resizebox{\hsize}{!}{\includegraphics{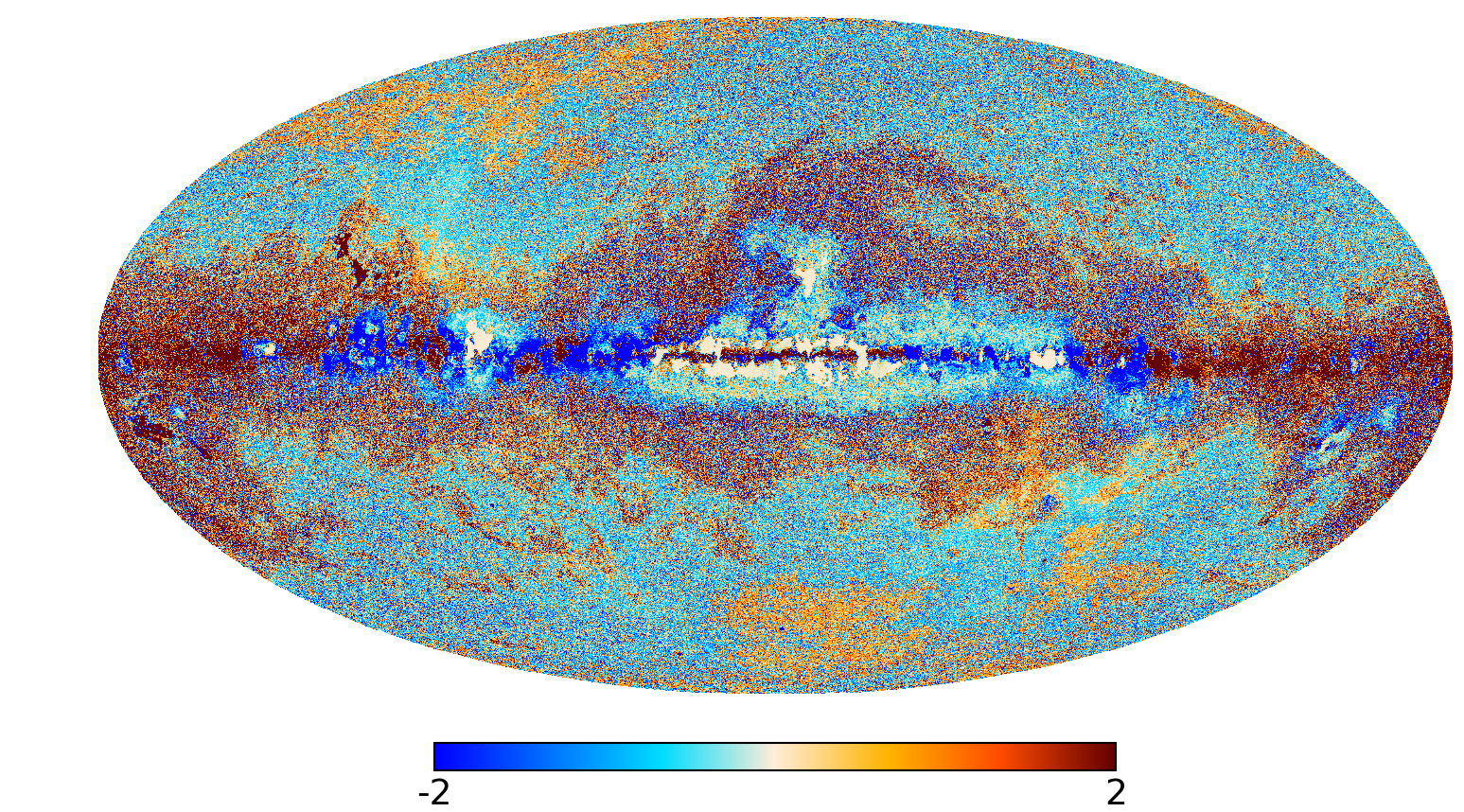}}
\caption{23 clusters}
\end{subfigure}
\caption{\label{Fig:part}The change in FC-ILC solution when changing the
  number of clusters the map is divided into. All maps are difference from
  the solution obtained with 13 clusters.  The results are shown for the FFP6
simulation. Similar results are obtained also for the real sky data. The
difference map has been degraded from the original $\Nside=2048$ to
$\Nside=512$ for this figure. The color scale ranges from $-1~{\rm \mu K}$ to
$1~{\rm \mu K}$ for the difference maps for 11 and 15 clusters  and larger
ranges for the other two difference maps.}
\end{figure}

We show the effect of changing the number of pixel clusters (including two fixed
clusters covering the dirtiest portions of the sky) in
Fig. \ref{Fig:part}. We show the difference in FC-ILC solution for
different choices of number of clusters  from the solution for 13
clusters. The difference in solution is less than $1 ~{\rm \mu K}$ if we
change the number of pixel clusters to 11 or 15 but we see a slightly larger
difference if we choose the number of clusters to be too small (5) or too
large (23). Choosing the number of clusters around 13 gives the best
results as gauged by comparison of the input and output power spectra and
we choose to partition the map into 13 clusters (including 2 contaminated or
bad clusters with extreme values of the foreground measure $m$.

\end{document}